\documentclass[pre,aps,floats,twocolumn,superscriptaddress,floatfix,nofootinbib]
{revtex4}
\usepackage{graphicx,amssymb,amsmath,ifthen,rotating}
\usepackage[active]{srcltx}
\usepackage{color}
\begin{document}
\title{Self-gravitating clusters of Bose-Einstein gas with planar, cylindrical, or spherical symmetry: gaseous density profiles and onset of condensation}  
\author{Michael Kirejczyk}
\affiliation{
  Department of Physics,
  University of Rhode Island,
  Kingston RI 02881, USA}
  \author{Gerhard M{\"u}ller}
\affiliation{
  Department of Physics,
  University of Rhode Island,
  Kingston RI 02881, USA}
  \author{Pierre-Henri Chavanis}
\affiliation{
Laboratoire de Physique Th\'eorique, CNRS, 
Universit{\'e} Paul Sabatier, 
31062 Toulouse, France}

\begin{abstract}
We calculate density profiles for self-gravitating clusters of an ideal Bose-Einstein gas with nonrelativistic energy-momentum relation and macroscopic mass at thermal equilibrium.
Our study includes clusters with planar symmetry in dimensions $\mathcal{D}=1,2,3$, clusters with cylindrical symmetry in $\mathcal{D}=2,3$, and clusters with spherical symmetry in $\mathcal{D}=3$.
Wall confinement is imposed where needed to prevent escape.
The length scale and energy scale in use for the gaseous phase render density profiles for gaseous macrostates independent of total mass. 
Density profiles for mixed-phase macrostates have a condensed core surrounded by a gaseous halo.
The spatial extension of the core is negligibly small on the length scale tailored for the halo.
The mechanical stability conditions as evident in caloric curves permit multiple macrostates to coexist.
Their status regarding thermal equilibrium is examined by a comparison of free energies.
The onset of condensation takes place at a nonzero temperature in all cases.
The critical singularities and the nature of the phase transition vary with the symmetry of the cluster and the dimensionality of the space.
\end{abstract}
\maketitle

%
\section{Introduction}\label{sec:intro}
%
In 1907, when Emden's work on \emph{Gaskugeln} was published \cite{emden}, the atomic nature of matter had barely escaped from controversy.
The thermodynamics of self-gravitating systems has not ceased to fascinate ever since \cite{paddy,found,ijmpb}. 
The pioneering work of Antonov \cite{antonov} considered an isolated system of nonrelativistic classical particles in gravitational interaction with given total mass $M$ and energy $E$, a reasonable starting point to model stellar systems including globular clusters. 
Microcanonical equilibrium states are obtained by maximizing the Boltzmann entropy $S$ at fixed mass $M$ and energy $E$ by using the notion of most probable macrostate \cite{ogorodnikov}. 
This leads to the mean-field Boltzmann distribution which is self-consistently coupled to the Poisson equation.

The Boltzmann-Poisson equation was previously studied in the
context of isothermal stars \cite{emden,chandrabook2}. It can be
reduced to the Emden equation and requires a numerical analysis. 
Antonov \cite{antonov} introduced wall confinement (at radius $R$),
which is necessary to stabilize finite-mass solutions of the Emden equation against escape. 

Antonov \cite{antonov} also introduced the widely used density contrast, 
$\mathcal{R}\doteq\rho_0/\rho(R)$ as a key parameter for wall-confined clusters
and determined, by analyzing second variations of the entropy $S$, the condition $\mathcal{R}<709$ for the stability of solutions of the Emden equation against a collapse.
This line of work was extended by Lynden-Bell and Wood \cite{lbw}, who calculated the energy $E(\mathcal{R})$ of solutions and analyzed the stability of clusters using criteria based on the Poincar\'e turning point argument \cite{poincare}.
They found stability for $E>E_c=E(\mathcal{R}_c)=-0.335\, GM^2/R$ and
$\mathcal{R}<\mathcal{R}_c=709$, thus confirming Antonov's prediction. The
instability for $\mathcal{R}>\mathcal{R}_c$ named
\emph{gravothermal catastrophe} is caused by  the
negative specific heat of the central region of the system.

By considering also the canonical ensemble, where equilibrium states are
associated with 
a minimum of the free energy $F=E-TS$,
Lynden-Bell and Wood
\cite{lbw} encountered an instance of ensemble inequivalence, a now well investigated peculiarity of thermodynamic systems with long-range interactions \cite{campabook}.
They specifically found stability for temperatures $T>T_c=T(\mathcal{R}'_c)=0.397\, GMm/k_BR$ and
$\mathcal{R}<\mathcal{R}'_c=32.1$, thus confirming earlier results of Emden \cite{emden}. 
Note that $\mathcal{R}'_c<\mathcal{R}_c$.
Similar results were obtained independently by Thirring \cite{thirring}.

A more general method for determining the stability of macrostates, in extension of Poincar\'e's theory, was developed by Katz \cite{katzpoincare1}. 
Its predictions are inferred from the topology of caloric curves: inverse temperature versus negative energy, $\beta(-E)$.
In the microcanonical ensemble, instabilities occur at turning points of energy, whereas in the canonical ensemble they occur at turning points of temperature.
Stability is lost (gained) if the curve $\beta(-E)$ turns clockwise (counterclockwise).
This early work has since been built-on by many further studies
\cite{paddyapj,dvsc,dvs1,dvs2,aaiso,grand,katzokamoto,metastable}.

Clusters of lower symmetry are, effectively, of lower dimensionality if classical  statistics is applicable, which is the case if the gas is sufficiently dilute everywhere.
Self-gravitating gaseous filaments (sheets) are astrophysical representations of clusters with cylindrical (planar) symmetry and are effectively two-dimensional (one-dimensional).
Mathematically, it is straightforward to extend the dimensionality to
$\mathcal{D}\neq 3$. 
There is much to learn from the $\mathcal{D}$-dependence of self-gravitating classical gas clusters.

The thermodynamics of self-gravitating classical gas clusters in $\mathcal{D}=2$ was pioneered by Stodolkiewicz \cite{stodolkiewicz}, Ostriker \cite{ostriker}, Salzberg \cite{salzberg}, and Katz and Lynden-Bell \cite{klb}. 
The evidence showed that an equilibrium state exists for all energies $E$ in the microcanonical ensemble, but only for temperatures $T\ge T_c=GMm/4k_B$ in the canonical ensemble. 
The caloric curve $\beta(E)$ is monotonic, which implies mechanical stability.
When the wall confinement is gradually moved out to infinity,
it can be demonstrated (by use of the virial theorem \cite{virialD}) that the equilibrium states all coalesce at the same  temperature, $T=T_c=GMm/4k_B$.
This result was first shown by  Stodolkiewicz \cite{stodolkiewicz}, Ostriker \cite{ostriker}, and Salzberg \cite{salzberg}, but it is already implicit in the work of Chandrasekhar and Fermi \cite{cf}.
It can be extended beyond the mean field approximation $N\gg 1$, producing the
exact result $T_c^{\rm exact}=(G/4N)\sum_{i\neq j}m_im_j$ \cite{virialD}.
Interestingly, a similar result appears in the statistical mechanics of 2D point vortices and in the chemotaxis of bacterial populations (see the discussion in \cite{cmct,bppv}).
In the microcanonical ensemble, these equilibrium states pertain to different energies.
Classical gas clusters in $\mathcal{D}=2$ were further investigated in several 
studies  \cite{paddy2d, aly, ar, ap} with results that connect to
this work.

Work on self-gravitating classical gas clusters in $\mathcal{D}=1$ began with
Spitzer \cite{spitzer}, Camm \cite{camm}, Rybicki \cite{rybicki}, and Katz and
Lecar \cite{kl}. 
Stable and unique equilibrium macrostates exist for all energies or temperatures  with or without wall confinement. The (monotonic) caloric curve for the latter case, as inferred from the virial theorem \cite{virialD},  is $E=\frac{3}{2}Nk_B T$. 
There are no mechanical instabilities.
Rybicki's work  \cite{rybicki} went beyond the mean-field framework and produced
some exact results for self-gravitating systems in $\mathcal{D}=1$.

A systematic extension of these low-$\mathcal{D}$ studies to arbitrary values of the spatial dimension for wall-confined systems was carried out by Sire and Chavanis \cite{sc}. 
They identified marginal dimensionalities, which delimit regimes of qualitatively different behavior.
The effects of wall confinement 
in self-gravitating systems were investigated with greater detail by Chavanis 
\cite{cmct,bppv}. 
A more recent study focused on density profiles of a self-gravitating lattice  gas in $\mathcal{D}=1,2,3$ \cite{selgra} (see also \cite{fdps}). 
The lattice gas has a built-in short-range repulsion, which produces effects akin to those of the exclusion principle in fermionic quantum gases \cite{sgcfd}.

Quantum mechanics stabilizes self-gravitating clusters against gravitational collapse.
This is universally true for bosons and fermions in the nonelativistic regime \cite{ijmpb}, but here the focus is on bosons.
The concept of a boson star was born in work aiming to determine the ground
state of boson clusters in the framework of Newtonian gravity and general
relativity \cite{jetzer}.
In general relativity, a self-gravitating Bose-Einstein condensate (BEC) is described by the Klein-Gordon-Einstein equations.  

The resulting mass-radius relation indicates the absence of an equilibrium state if the mass exceeds the value $M_{\rm max}=0.633\, M_P^2/m$ \cite{kaup,rb} where $M_P=(\hbar c/G)^{1/2}$ is the Planck mass. 
At this point, the boson star is expected to collapse and form a black hole.
These results are similar to those obtained in the case of general 
relativistic fermion stars (e.g. neutron stars) whose  maximum mass is given by
$M_{\rm max}=0.384\, M_P^3/m^2$ \cite{ov}. Note the scaling $m^{-1}$ instead of
$m^{-2}$, which is due to the fact that boson stars
are stabilized by the Heisenberg uncertainty principle while fermion stars are
stabilized by the Pauli exclusion principle. As a result, for the same
particle mass $m$, with $m\ll M_P$, the maximum mass of noninteracting boson
stars is
much smaller than the maximum mass of fermion stars. 

A self-gravitating BEC in the framework of Newtonian gravity, described by the Schr\"odinger-Poisson equations, does not have compact support. 
The mass-radius relations used in such contexts introduce $R_{99}$, the radius that encloses 99\% of the total mass.
The result, $M=9.95\, \hbar^2/(Gm^2R_{99})$, emerging from several studies \cite{membrado,prd1,prd2} is similar to what has been well-known for nonrelativistic fermion stars (e.g. white dwarfs), but on different scales: $M=91.9\, \hbar^6/(G^3m^8R^3)$ \cite{chandrabook2}. 
There is yet no clear evidence for the existence of boson stars.
However, it has been proposed that the core of neutron stars might turn superfluid via a kind of neutron pairing into bosons, thus forming a relativistic BEC of sorts \cite{chavharko}. 

It has also been proposed that dark matter (DM) halos may be
made of ultralight 
bosons (axions) with a mass $m\sim 10^{-22}\, {\rm eV}/c^2$, named fuzzy dark
matter (FDM) \cite{hu}.
These ultralight particles are still hypothetical but they are not excluded by particle physics and are actively studied at present \cite{hui}. Another type of massive bosons that could constitute dark matter is the QCD axion, a pseudo-Nambu-Goldstone boson of the Peccei-Quinn \cite{pq} phase transition associated with a $U(1)$ symmetry that solves the strong charge parity (CP) problem of quantum chromodynamics (QCD).
However, its mass $m=10^{-4}\, {\rm eV}/c^2$ is larger so it yields smaller astrophysical structures called axion stars \cite{braatenrevue}. 
The large occupancies of axions permit such halos to be described by the Schr\"odinger-Poisson equations.
The analysis reveals
gravitational cooling and violent relaxation processes
\cite{seidel94,lb}.
Supporting evidence for these processes comes from  numerical simulations of the Schr\"odinger-Poisson equations \cite{ch2, ch3, schwabe, mocz, moczSV, veltmaat, moczprl, moczmnras, veltmaat2}.

The BEC core (often called soliton) has a size of the order of the de Broglie length $\lambda_{\rm dB}=h/(mv)\sim 1\, {\rm kpc}$. 
The surrounding halo results from quantum interferences of excited states. 
It has a profile similar to the  Navarro-Frenk-White profile obtained in
numerical simulations of classical cold dark matter (CDM) \cite{nfw}. 
Such a core-halo profile is reminiscent of those emerging from the Lynden-Bell statistical theory  of collisionless violent relaxation \cite{lb,csr}. 
In particular, an approximately isothermal halo can account for the flat rotation curves of the galaxies. 

Quantum statistics stabilizes matter against gravitational collapse at small
scales by producing a soliton core in replacement of a cusp such as obtained in
simulations of 
classical CDM models \cite{nfw}. 
This difference is significant because observations
\cite{observations} favor cores over cusps. 
Quantum statistics may be a way to solve the core-cusp problem of the classical
CDM modeling. 

Core-halo structures resulting from violent relaxation typically are mechanically stable, but do not represent thermal equilibrium states.
The halo is only approximately an isothermal distribution in the sense of Lynden-Bell \cite{lb,csr}. 
Its effective temperature $T_{\rm eff}$ has typical values far below criticality
of a BE gas with the same parameters, which raises interesting
questions regarding particle masses of bosonic DM addressed in this work.

Current assumptions about bosonic DM particle mass imply that the number of
bosons in a typical halo is gigantic: $N\sim 10^{99}$.
The Chandrasekhar relaxation time $t_{\rm relax}\sim (N/\ln N)t_D$, where $t_D\sim 10^8\, {\rm yrs}$, exceeds the age of the universe by far. 
If the particles were classical, the DM halo would be effectively collisionless.
The halo would not have had time to relax
towards a thermalized state.

However, quantum interferences cause the halo to have a granular structure 
with a correlation length $\sim \lambda_{\rm dB}$ \cite{ch2,ch3}. 
The granules are quasi-particles of effective mass $m_{\rm
eff}\sim\rho \lambda_{\rm dB}^3\sim 10^7\,
M_{\odot}$ \cite{hui}, much larger than the postulated typical DM boson mass
$m\sim 10^{-22}\, {\rm eV}/c^2$.
The number of particles in a bosonic DM halo is thus effectively reduced to 
$N_{\rm eff}\sim 10^5$,  
which is comparable to the number of stars in a typical globular cluster. 

Therefore, granular effects are important.
They induce a collisional evolution of the DM halo on an accelerated time scale, of 
the order of the Hubble time \cite{bft,bft2,chavlandau}. 
This evolution toward thermal equilibrium (at very low $T$) has the effect of triggering a condensation. 
The halo slowly condenses with the solitonic core progressively growing in mass
\cite{levkov2,egg}. 
However, a complete state of statistical equilibrium
 is reached on a very long time scale.

The work reported in this study is not aiming to answer specific open questions in current DM research or research on boson stars. 
It is centered in equilibrium statistical mechanics with possible applications in astrophysics.
Its focus is on the gravity driven condensation in clusters of different symmetry in spaces of different dimensionality. 
Theories of pure gas clusters and pure condensates operate on length scales whose ratio involves powers of the number of particles involved.
The huge disparity in length scale between a BEC and its gaseous halo is a challenge. 
The density profile of the gaseous halo is strongly influenced by the gravity of the BEC core, whose density profile, in turn, is strongly affected by the weight of the surrounding gas.

The analysis carried out in this work, which focuses on the gaseous halo, introduces a provisional BEC in the form of a reference state that represents a core of high and uniform density. 
It will be argued that this scheme yields an accurate account of the onset of condensation and the nature of the phase transitions.
A separate study will have to be carried out on a much contracted length scale to analyze the deviation of the BEC density profile from a uniform shape under the weight of the gaseous halo determined in this study.

Our work builds on previous studies in the same line, which are few in number.
The work of Ingrosso and Ruffini \cite{ir} stays within the framework of Newtonian gravity, whereas the work of Bili\'c and Nikoli\'c \cite{bn} widens the framework to general relativity. 
Both studies yield important results, which serve as benchmarks in our work.
Our study is limited to nonrelativistic bosons, but considers clusters of different symmetry and spaces of different dimensionality.

We begin by establishing conditions for thermal equilibrium and mechanical
stability, by introducing useful energy and length scales, and by deriving
free energy integral expressions that also cover two-phase macrostates
(Sec.~\ref{sec:fund}). 
We continue with a detailed account of the density profiles and the phase behavior of planar clusters in one, two, and three dimensions (Sec.~\ref{sec:planar}), of cylindrical clusters in two and three dimensions (\ref{sec:cylind}), and of spherical clusters in three dimensions (\ref{sec:sphere}). 

%
\section{Fundamentals}\label{sec:fund}
%
The fundamental ingredients to this work include 
(i) the conditions of thermal equilibrium and mechanical equilibrium for gas
clusters and mixed phase clusters consisting of a BEC core surrounded by a
gaseous halo, 
(ii) the choice of a length scale adequate for the description of gaseous density profiles at all temperatures including the Maxwell-Boltzmann (MB) limit, and 
(iii) the establishment of a free energy expression on an energy scale that
covers all scenarios of interest.
Implied in this scheme is the validity of mean-field assumptions owing to the 
long-range nature of the gravitational interaction, and supported by studies
dedicated to this question \cite{ht}.

\subsection{Thermal equilibrium}\label{sec:therm-eq}
The equation of state (EOS) for the nonrelativistic BE gas in $\mathcal{D}$
dimensions of particles with mass $m$ is implicit in the fundamental
thermodynamic relations \cite{ll,reichl,schwabl},
\begin{subequations}\label{eq:1}
\begin{align}\label{eq:1a}
& \frac{p\lambda_T^\mathcal{D}}{k_\mathrm{B}T}=g_s\,g_{\mathcal{D}/2+1}(z), \\ 
\label{eq:1b}
& \rho_\mathrm{v}\lambda_T^\mathcal{D}=g_s\,g_{\mathcal{D}/2}(z), \\ 
\label{eq:1c}
& u_\mathrm{v}\lambda_T^\mathcal{D}=\frac{\mathcal{D}}{2}k_\mathrm{B}Tg_s\,g_{\mathcal{D}/2+1}(z), 
\end{align}
\end{subequations}
where $p$ is the pressure,  $\rho_\mathrm{v}\doteq N/V$ the
particle
density, $u_\mathrm{v}\doteq U/V$ the kinetic-energy density, $z$ the
fugacity, $g_s$ the spin degeneracy,
\begin{equation}\label{eq:2}
\lambda_T=\sqrt{\frac{h^2}{2\pi mk_\mathrm{B}T}},
\quad \beta\doteq\frac{1}{k_\mathrm{B}T},
\end{equation}
the de Broglie thermal wavelength, 
and 
\begin{equation}\label{eq:3}
g_n(z)\doteq\frac{1}{\Gamma(n)}\int_0^\infty\frac{dx\,x^{n-1}}{z^{-1}e^x-1}
=\sum_{l=1}^\infty\frac{z^l}{l^n},\quad 0\leq z\leq1
\end{equation}
the (polylogarithmic) BE functions. 
Equations~(\ref{eq:1}) are
inferred from the grand partition function $Z$
via the grand potential $\Omega(T,V,\mu)=-k_\mathrm{B}T\ln Z=-pV$,
whose natural independent variables are temperature $T$, volume $V$,
and chemical potential $\mu=k_\mathrm{B}T\ln z$. 
The entropy density,
$s_\mathrm{v}\doteq S/V$, can be inferred from (\ref{eq:1}) via Euler's
equation, $U=TS-pV+\mu N$:
\begin{equation}\label{eq:125} 
\frac{S\lambda_T^\mathcal{D}}{g_sVk_\mathrm{B}}=
\left(\frac{\mathcal{D}}{2}+1\right)g_{\mathcal{D}/2+1}(z)
-\ln z\,g_{\mathcal{D}/2}(z).
\end{equation}
Expressions~(\ref{eq:1}) and (\ref{eq:125}) are taken to hold locally in the gaseous part of a BE cluster, over distances that are short on the length scale to be introduced for the characterization of gaseous density profiles at all temperatures and in all spatial dimensions.

\subsection{Mechanical equilibrium}\label{sec:mech-eq}
In a cluster of self-gravitating gas at equilibrium, the temperature $T$ is
uniform, but the pressure $p$ and the particle density $\rho_\mathrm{v}$ acquire
profiles to satisfy mechanical stability. 
It is expedient to
introduce a second discrete parameter,
$\mathcal{D}_\sigma$, for the purpose of characterizing the symmetry
of the cluster under scrutiny. It will naturally appear in expressions
which are valid for clusters of different symmetry. 

We consider clusters with
planar symmetry $(\mathcal{D}_\sigma=1)$, cylindrical symmetry
$(\mathcal{D}_\sigma=2)$, and spherical symmetry $(\mathcal{D}_\sigma=3)$.
All profiles are functions of the distance $r$ from the center of the cluster.
For $\mathcal{D}_\sigma=1$, the center is a point, a line, or a plane in $\mathcal{D}=1,2,3$, respectively. 
For $\mathcal{D}_\sigma=2$, the center is a point or a line in $\mathcal{D}=2,3$, respectively. 
For $\mathcal{D}_\sigma=3$, the center is a point (in $\mathcal{D}=3$). 
We thus write $\rho_\mathrm{v}(r)$, $p(r)$, $z(r)$, and $\mu(r)$ for the radial profiles of particle density, pressure, fugacity, and chemical potential, respectively.

The total number of particles in a finite cluster is obtained from the density profile via the integral,
\begin{equation}\label{eq:4} 
N=L^{\mathcal{D}-\mathcal{D}_\sigma}
\int_0^R dr \, \mathcal{A}_{\mathcal{D}_\sigma} 
r^{\mathcal{D}_\sigma-1}\rho_\mathrm{v}(r),
\end{equation}
where $R$ is the radius of the confining wall, $L$ the length of the cylinder or the sides of the plane, and
\begin{equation}\label{eq:5} 
\mathcal{A}_\mathcal{D}\doteq\frac{2\pi^{\mathcal{D}/2}}{\Gamma(\mathcal{D}/2)}
=\left\{\begin{array}{ll}
2 &: \mathcal{D}=1, \rule[-2mm]{0mm}{3mm}\\ 2\pi &: \mathcal{D}=2, \rule[-2mm]{0mm}{3mm}\\ 4\pi &: \mathcal{D}=3, \rule[-2mm]{0mm}{3mm}
\end{array} \right.
\end{equation}
is the surface area of the $\mathcal{D}$-dimensional unit sphere.
The condition $L\gg R$ guarantees that deviations from the symmetry assumed to hold are negligible.
The mechanical equilibrium is governed by the equation of motion (EOM), here expressing hydrostatic equilibrium,
\begin{equation}\label{eq:6} 
\frac{d}{dr}p(r)=m\rho_\mathrm{v}(r)g(r).
\end{equation}
The gravitational field is inferred from Gauss's law: 
\begin{equation}\label{eq:8} 
g(r)=-\frac{\mathcal{A}_\mathcal{D}G_\mathcal{D}m}{r^{\mathcal{D}_\sigma-1}}
\int_0^r dr'r'^{\mathcal{D}_\sigma-1}\rho_\mathrm{v}(r').
\end{equation}

\subsection{Fugacity and chemical potential}\label{sec:fug-chem}
The primary profile to be calculated will be $z(r)$ for all cases with the exception of critical macrostates, where the direct calculation $\mu(r)=k_\mathrm{B}T\ln z$ offers some advantages.
Carrying out the derivative of $p(r)$ using (\ref{eq:1}) and the recurrence
relation, $zg_n'(z)=g_{n-1}(z)$, yields
\begin{equation}\label{eq:9} 
p'(r)=k_\mathrm{B}T\,\frac{z'(r)}{z(r)}\,\rho_\mathrm{v}(r).
\end{equation}
Equation (\ref{eq:9}) is more general than Eq.~(\ref{eq:1a}) for the characterization of pressure profiles. The latter is an integral version of the former, restricted to cases where $z(r)$ is a continuous function. 
It converts (\ref{eq:6}) into
\begin{equation}\label{eq:9aa} 
k_\mathrm{B}T\,\frac{z'(r)}{z(r)}=mg(r),\quad g(r)\doteq-\frac{d\mathcal{U}}{dr},
\end{equation}
from which a familiar relation between fugacity $z(r)$ and gravitational potential $\mathcal{U}(r)$ follows upon integration:
\begin{equation}\label{eq:235}
z(r)=z_0\,e^{-\beta m[\mathcal{U}(r)-\mathcal{U}_0]}.
\end{equation}

The derivation of a differential equation for $z(r)$ combines (\ref{eq:8}) and (\ref{eq:9aa}) into
\begin{equation}\label{eq:10} 
\frac{z'(r)}{z(r)}r^{\mathcal{D}_\sigma-1}
=-\frac{\mathcal{A_D}G_\mathcal{D}m^2}{k_\mathrm{B}T}
\int_0^r dr'r'^{\mathcal{D}_\sigma-1}\rho_\mathrm{v}(r'),
\end{equation}
which, upon differentiation and use of (\ref{eq:1b}), yields the following ODE for the fugacity profile:
\begin{equation}\label{eq:11} 
\frac{z''}{z}+\frac{\mathcal{D}_\sigma-1}{r}\frac{z'}{z}-\left(\frac{z'}{z}\right)^2
+\frac{\mathcal{A_D}G_\mathcal{D}m^2}{\lambda_T^\mathcal{D}k_\mathrm{B}T}g_sg_{\mathcal{D}/2}(z)=0.
\end{equation}
The profiles for pressure and density follow directly.

For (thermodynamically) open BE gas clusters of finite or infinite mass, the boundary conditions are
\begin{equation}\label{eq:12} 
z'(0)=0,\quad 0<z(0)=z_0\leq1,
\end{equation}
with the (average) total mass, $Nm$, provided it is finite, inferred from (\ref{eq:4}).
Closed systems of finite mass $Nm$ may not exist without confinement.
For systems with ${\mathcal{D}_\sigma<\mathcal{D}}$, it is useful to rescale the number of particles:
\begin{equation}\label{eq:13}
\tilde{N}=\frac{N}{L^{\mathcal{D}-\mathcal{D}_\sigma}}.
\end{equation}
The second boundary condition (\ref{eq:12}) must then be replaced by the integral condition (\ref{eq:4}) converted into
\begin{equation}\label{eq:14}
\frac{g_s\mathcal{A}_{\mathcal{D}_\sigma}}{\tilde{N}\lambda_{T}^{\mathcal{D}}}
\int_0^Rdr\,r^{\mathcal{D}_\sigma-1}g_{\mathcal{D}/2}(z)=1.
\end{equation}
In macrostates consisting of a BEC core surrounded by a gaseous halo, the gas is still described by the ODE (\ref{eq:11}), but with modified boundary conditions (see Sec.~\ref{sec:free-ener}).

\subsection{Scaling convention}\label{sec:le-scale}
The physics of self-gravitating BE gas clusters unfolds on a characteristic length scale and a characteristic energy (or temperature) scale.
The scaling convention adopted in this work captures both scales via the thermal wavelength and a dimensional analysis of the ODE (\ref{eq:11}).
We write, 
\begin{equation}\label{eq:15} 
\hat{r}\doteq\frac{r}{r_\mathrm{s}},\quad \hat{T}\doteq\frac{T}{T_\mathrm{s}},
\end{equation}
with $r_\mathrm{s}$ and $T_\mathrm{s}$ from
\begin{equation}\label{eq:16} 
\tilde{N}\lambda_{T_\mathrm{s}}^{\mathcal{D}}
=\frac{\mathcal{A}_{\mathcal{D}_\sigma}}{\mathcal{D}_\sigma}\,r_\mathrm{s}^{\mathcal{D}_\sigma}
=\left\{\begin{array}{ll}
2r_\mathrm{s} &: \mathcal{D}_\sigma=1, \rule[-2mm]{0mm}{5mm}\\ \pi r_\mathrm{s}^2 &: \mathcal{D}_\sigma=2, \rule[-2mm]{0mm}{5mm}\\ {\displaystyle \frac{4\pi}{3}r_\mathrm{s}^3} &: \mathcal{D}_\sigma=3, \rule[-2mm]{0mm}{5mm}
\end{array} \right.
\end{equation}
\begin{equation}\label{eq:17} 
\frac{1}{r_\mathrm{s}^2}=\frac{1}{2\mathcal{D}_\sigma}\frac{\mathcal{A_D}G_\mathcal{D}m^2}
{\lambda_{T_\mathrm{s}}^{\mathcal{D}}k_\mathrm{B}T_\mathrm{s}}.
\end{equation}
Equation (\ref{eq:16}) attributes a volume $\lambda_{T_\mathrm{s}}^{\mathcal{D}}$ to each particle. 
$\tilde{N}$ such particles form a compact cluster of radius $r_\mathrm{s}$ and given symmetry.
Equations (\ref{eq:16}) and (\ref{eq:17}) determine $r_\mathrm{s}$ and
$T_\mathrm{s}$ as functions of particle mass $m$ and total mass
$\tilde{m}_\mathrm{tot}$.
In cases with $\mathcal{D}_\sigma<D$, the quantity
$\tilde{m}_\mathrm{tot}=\tilde{N}m$ is the total mass per unit length or unit
area in the directions of translational symmetry.
It is noteworthy that the scales $r_\mathrm{s}$ and $T_\mathrm{s}$ are equally useful for Fermi-Dirac (FD) clusters \cite{sgcfd}. 
The distinct dependences on particle mass and on total mass spelled out in 
Appendix~\ref{sec:appb} are an attribute of potential importance in DM
research.

With the dimensionless variables thus defined, we set $\hat{z}(\hat{r})\doteq z(r)$ and convert the ODE (\ref{eq:11}) into
\begin{equation}\label{eq:18} 
\frac{\hat{z}''}{\hat{z}}+\frac{\mathcal{D}_\sigma-1}{\hat{r}}\frac{\hat{z}'}{\hat{z}}
-\left(\frac{\hat{z}'}{\hat{z}}\right)^2
+\frac{2\mathcal{D}_\sigma}{\hat{T}^{1-\mathcal{D}/2}}g_sg_{\mathcal{D}/2}(\hat{z})=0.
\end{equation}
The initial conditions are $\hat{z}'(0)=0$ and $\hat{z}(0)$ from
\begin{equation}\label{eq:20}
\mathcal{D}_\sigma \int_0^{\hat{R}} d\hat{r}\,\hat{r}^{\mathcal{D}_\sigma-1}
\rho(\hat{r})=1,
\end{equation}
a rescaled Eq.~(\ref{eq:14}).
Henceforth we shall use the dimensionless density,
\begin{equation}\label{eq:19}
\rho(\hat{r})\doteq \lambda_{T_\mathrm{s}}^\mathcal{D}\rho_\mathrm{v}(\hat{r})
=g_s\,\hat{T}^{\mathcal{D}/2}g_{\mathcal{D}/2}(\hat{z}).
\end{equation}

All results expressed with these scales are independent of the number of particles (or total mass) provided it is macroscopic and relativistic effects are negligible \cite{sgcrfd}.
If a coexisting condensate is present at the core of the BE gas cluster, it must be described on a different length scale, one that is tiny in units of $r_\mathrm{s}$ (see below).

\subsection{Maxwell-Boltzmann limit}\label{sec:MB-limit}
When the gas is dilute throughout the cluster, which implies that $\hat{z}\ll1$ everywhere, we can simplify (\ref{eq:19}) into
\begin{equation}\label{eq:216}
\rho(\hat{r})\leadsto g_s\hat{T}^{\mathcal{D}/2}\hat{z}.
\end{equation}
Equation~(\ref{eq:18}) for the fugacity thus simplifies into an ODE for the density,
\begin{equation}\label{eq:21} 
\frac{\rho''}{\rho}+\frac{\mathcal{D}_\sigma-1}{\hat{r}}\frac{\rho'}{\rho}-\left(\frac{\rho'}{\rho}\right)^2
+\frac{2\mathcal{D}_\sigma}{\hat{T}}\rho=0,
\end{equation}
with boundary conditions, $\rho'(0)=0$, $\rho(0)=\rho_0$, and (\ref{eq:20}).
Equation~(\ref{eq:21}) is equivalent to the Emden equation \cite{chandrabook2} and well known to be characteristic of the MB gas.
It only depends on $\mathcal{D}_\sigma$, the symmetry of the cluster.
In Ref.~\cite{selgra} the MB gas emerged (with $\mathcal{D}_\sigma=\mathcal{D}$) in the dilute limit of the ideal lattice gas.
The scaled variables remain the same, but the length scale is different.
The volume unit $\lambda_{T_\mathrm{s}}^\mathcal{D}$ replaces
the lattice-gas cell volume $V_\mathrm{c}$.

The ODE (\ref{eq:21}) for the MB limit of the BE gas is invariant under the following scale transformation for arbitrary (dimensionless) $\hat{r}_\mathrm{t}>0$:
\begin{equation}\label{eq:219}
\tilde{r}\doteq\frac{\hat{r}}{\hat{r}_\mathrm{t}},\quad 
\tilde{\rho}\doteq \hat{r}_\mathrm{t}^{\mathcal{D}_\sigma}\rho,\quad 
\tilde{T}\doteq \hat{r}_\mathrm{t}^{\mathcal{D}_\sigma-2}\hat{T},\quad
\tilde{R}\doteq\frac{\hat{R}}{\hat{r}_\mathrm{t}}.
\end{equation}
When we set $\hat{r}_\mathrm{t}=\hat{R}$, the scale transformation produces a universal MB density profile that covers any radius $\hat{R}$ of confinement \cite{selgra}:
\begin{subequations}\label{eq:217}
\begin{eqnarray}\label{eq:217a}
\tilde{\rho}''+\frac{\mathcal{D}_\sigma-1}{\tilde{r}}\frac{\tilde{\rho}'}{\tilde{\rho}}
-\left(\frac{\tilde{\rho}'}{\tilde{\rho}}\right)^2
+\frac{2\mathcal{D}_\sigma}{\tilde{T}}\,\tilde{\rho}=0,
\end{eqnarray}
\begin{equation}\label{eq:217b}
\tilde{\rho}'(0)=0,\quad 
\mathcal{D}_\sigma\int_0^1 d\tilde{r}\,\tilde{r}^{\mathcal{D}_\sigma-1}\tilde{\rho}(\tilde{r})=1.
\end{equation}
\end{subequations}
This universality does not hold for BE or FD gas clusters in general.
We find qualitative changes in phase behavior of BE clusters, when the radius
of
confinement is varied.
Similar evidence is reported in \cite{sgcfd} for FD clusters.

\subsection{Free energy}\label{sec:free-ener}
Caloric curves and free energy comparisons are common tools for determining the
stability status of competing solutions of the ODE (\ref{eq:18}).
Here we develop the ingredients to these tools: entropy $S$, internal energy $E$, and Helmholtz free energy $F$.
We can write
\begin{equation}\label{eq:222}
F=E-TS=U+W-TS,
\end{equation}
where $U$ is the kinetic energy and $W$ the (gravitational) potential energy.
Using the kinetic energy density (\ref{eq:1c}) and the entropy density
(\ref{eq:125}), we can write the scaled integral expressions,
\begin{equation}\label{eq:223}
\hat{U}\doteq\frac{U}{Nk_\mathrm{B}T_\mathrm{s}}=
\mathcal{D}_\sigma g_s\hat{T}^{\mathcal{D}/2+1}\frac{\mathcal{D}}{2}
\int_0^{\hat{R}} d\hat{r}\,\hat{r}^{\mathcal{D}_\sigma-1}
g_{\mathcal{D}/2+1}(\hat{z}),
\end{equation}
\begin{align}\label{eq:224}
 \hat{S}\doteq &\frac{S}{Nk_\mathrm{B}}=
\mathcal{D}_\sigma g_s\hat{T}^{\mathcal{D}/2}
\int_0^{\hat{R}} d\hat{r}\,\hat{r}^{\mathcal{D}_\sigma-1} \nonumber \\
& \hspace{5mm}\times \left[\left(\frac{\mathcal{D}}{2}+1\right)g_{\mathcal{D}/2+1}(\hat{z})
-\ln\hat{z}\,g_{\mathcal{D}/2}(\hat{z})\right],
\end{align}
\begin{align}\label{eq:225}
\hat{U}-&\hat{T}\hat{S} =
-\mathcal{D}_\sigma g_s\hat{T}^{\mathcal{D}/2+1} \nonumber \\
& \times\int_0^{\hat{R}} d\hat{r}\,\hat{r}^{\mathcal{D}_\sigma-1}
\Big[g_{\mathcal{D}/2+1}(\hat{z})-\ln\hat{z}\,g_{\mathcal{D}/2}(\hat{z})\Big].
\end{align}

The calculation of $W$ faces issues related to the different length scales appropriate for the BEC core and the gaseous halo of mixed-state density profiles.
They will be addressed in Sec.~\ref{sec:BEC-rad}, once we have developed a working expression for the gravitational potential energy pertaining to clusters of different symmetry.
Our construction of $W$ adapts a scheme previously developed for the self-gravitating lattice gas \cite{selgra}.

We use a reference state which represents a core of radius $\hat{r}_\mathrm{c}$ and uniform density, whose value in our scaling convention becomes $\rho_\mathrm{c}=\hat{r}_\mathrm{c}^{-\mathcal{D}_\sigma}$.
This reference state is not meant to represent a physical state.
It is a placeholder for the BEC.
Its uniform density is a mere convenience.
We present expressions of $W$ for gaseous macrostates and mixed macrostates consisting of a gaseous halo surrounding a core of radius $0\leq\hat{r}_\mathrm{b}\leq\hat{r}_\mathrm{c}$ and uniform density $\rho_\mathrm{c}$, higher than the maximum gas density. 

\subsubsection{$\mathcal{D}_\sigma=1$}\label{sec:Ds=1}
For clusters with planar symmetry, the radius of the core is proportional to its fraction of total mass:
\begin{equation}\label{eq:239} 
\frac{\tilde{m}_\mathrm{b}}{\tilde{m}_\mathrm{tot}}=
\frac{\hat{r}_\mathrm{b}}{\hat{r}_\mathrm{c}}.
\end{equation}
The potential energy for a core-halo macrostate becomes
\begin{subequations}\label{eq:244}
\begin{align}\label{eq:244a} 
&\hat{W} \doteq \frac{W}{Nk_\mathrm{B}T_\mathrm{s}}
=2\hat{r}_\mathrm{c}^{-1}\int_{\hat{r}_\mathrm{b}}^{\hat{R}} d\hat{r}_2
\big[\hat{r}_2\hat{r}_1-\hat{r}_1^2\big]\rho(\hat{r}_2), \\ \label{eq:244b} 
&\hat{r}_1=\hat{r}_\mathrm{b}+\hat{r}_\mathrm{c}
\int_{\hat{r}_\mathrm{b}}^{\hat{r}_2}d\hat{r}\rho(\hat{r}),
\end{align}
\end{subequations}
where the density $\rho(\hat{r})$ of the gaseous halo is the solution of the ODE (\ref{eq:18}) with boundary conditions,
\begin{equation}\label{eq:242}
 \hat{z}'(\hat{r}_\mathrm{b})=
 -\frac{2}{\hat{T}}\frac{\hat{r}_\mathrm{b}}{\hat{r}_\mathrm{c}}\,
 \hat{z}(\hat{r}_\mathrm{b}),\quad \int_{\hat{r}_\mathrm{b}}^{\hat{R}}d\hat{r}\,\rho(\hat{r})
=1-\frac{\tilde{m}_\mathrm{b}}{\tilde{m}_\mathrm{tot}}.
\end{equation}
In the limit $\tilde{m}_\mathrm{b}\to0$ (gaseous macrostate), expression (\ref{eq:244}) simplifies into
\begin{align}\label{eq:92} 
\hat{W} 
=2\int_0^{\hat{R}} d\hat{r}_2\hat{r}_2\sigma_1(\hat{r}_2)\rho(\hat{r}_2)
-\frac{2}{3}\hat{r}_\mathrm{c},
\end{align}
\begin{equation}\label{eq:93} 
\sigma_1(\hat{r})
=\int_0^{\hat{r}} d\hat{r}'\rho(\hat{r}')
=-\frac{\hat{T}}{2}\frac{\hat{z}'(\hat{r})}{\hat{z}(\hat{r})}.
\end{equation}
The last expression of (\ref{eq:93}) effectively reduces the double integral in (\ref{eq:92}) into a single integral.
The reference state affects $W$ only as an additive constant.

\subsubsection{$\mathcal{D}_\sigma=2$}\label{sec:Ds=2}
In clusters with cylindrical symmetry, the mass of the core grows quadratically with the core radius,

\begin{equation}\label{eq:260} 
 \frac{\tilde{m}_\mathrm{b}}{\tilde{m}_\mathrm{tot}}
 =\left(\frac{\hat{r}_\mathrm{b}}{\hat{r}_\mathrm{c}}\right)^2.
\end{equation}
The potential energy of a core-halo state now reads,
\begin{subequations}\label{eq:264} 
\begin{align}\label{eq:264a} 
&\hat{W} 
=4\hat{r}_\mathrm{c}^{-2}\int_{\hat{r}_\mathrm{b}}^{\hat{R}} d\hat{r}_2
\hat{r}_2\hat{r}_1^2\rho(\hat{r}_2)\ln\left(\frac{\hat{r}_2}{\hat{r}_1}\right), 
\\ \label{eq:264b} 
&\hat{r}_1^2=\hat{r}_\mathrm{b}^2+2\hat{r}_\mathrm{c}^2
\int_{\hat{r}_\mathrm{b}}^{\hat{r}_2}d\hat{r}\,\hat{r}\rho(\hat{r}),
\end{align}
\end{subequations}
where the gas density inferred from (\ref{eq:18}) calls for the boundary conditions,
\begin{equation}\label{eq:262}
 \hat{z}'(\hat{r}_\mathrm{b})=
 -\frac{2}{\hat{T}}\frac{\hat{r}_\mathrm{b}}{\hat{r}_\mathrm{c}^2}\,
 \hat{z}(\hat{r}_\mathrm{b}), \quad 2\int_{\hat{r}_\mathrm{b}}^{\hat{R}}d\hat{r}\,\hat{r}\rho(\hat{r})
=1-\frac{\tilde{m}_\mathrm{b}}{\tilde{m}_\mathrm{tot}}.
\end{equation}
The simplified expression for the pure gas state is again effectively reducible to a single integral,
\begin{align}\label{eq:257}
\hat{W} =4\int_0^{\hat{R}} d\hat{r}_2\,\hat{r}_2\,\sigma_2(\hat{r}_2)\,\rho(\hat{r}_2)\ln\left(\frac{\hat{r}_2}{\sqrt{\sigma_2}}\right)
 -\ln\hat{r}_\mathrm{c}, 
\end{align}
\begin{equation}\label{eq:256}
\sigma_2(\hat{r})= 2\int_0^{\hat{r}}d\hat{r}'\,\hat{r}'\rho(\hat{r}')
=-\frac{\hat{T}}{2}\,\hat{r}\,\frac{\hat{z}'(\hat{r})}{\hat{z}(\hat{r})}.
\end{equation}

\subsubsection{$\mathcal{D}_\sigma=3$}\label{sec:Ds=3}
Systematic trends become apparent as we present the case of spherical symmetry. 
The mass-radius relation of the BEC is now cubic:
\begin{equation}\label{eq:280} 
 \frac{\tilde{m}_\mathrm{b}}{\tilde{m}_\mathrm{tot}}
 =\left(\frac{\hat{r}_\mathrm{b}}{\hat{r}_\mathrm{c}}\right)^3.
\end{equation}
The potential energy of a core-halo state returns to power-laws as seen for $\mathcal{D}_\sigma=1$:
\begin{subequations}
\begin{align}\label{eq:284a} 
&\hat{W} 
=\frac{6}{\hat{r}_\mathrm{c}^3}\int_{\hat{r}_\mathrm{b}}^{\hat{R}} d\hat{r}_2
\big[\hat{r}_2^2\hat{r}_1^2-\hat{r}_2\hat{r}_1^3\big]\rho(\hat{r}_2), 
\\ \label{eq:284b} 
&\hat{r}_1^3=\hat{r}_\mathrm{b}^3+3\hat{r}_\mathrm{c}^3
\int_{\hat{r}_\mathrm{b}}^{\hat{r}_2}d\hat{r}\,\hat{r}^2\rho(\hat{r}),
\end{align}
\end{subequations}
where the boundary conditions for (\ref{eq:18}) are 
\begin{equation}\label{eq:282}
 \hat{z}'(\hat{r}_\mathrm{b})=
 -\frac{2}{\hat{T}}\frac{\hat{r}_\mathrm{b}}{\hat{r}_\mathrm{c}^3}\,
 \hat{z}(\hat{r}_\mathrm{b}),\quad \!\!
 3\int_{\hat{r}_\mathrm{b}}^{\hat{R}}d\hat{r}\,\hat{r}^2\rho(\hat{r})
=1- \frac{\tilde{m}_\mathrm{b}}{\tilde{m}_\mathrm{tot}}.
\end{equation}
The simplified expressions for gaseous macrostates read:
\begin{align}\label{eq:274} 
\hat{W} =-6\int_0^{\hat{R}} d\hat{r}_2\hat{r}_2\sigma_3(\hat{r}_2)\rho(\hat{r}_2)
+\frac{6}{5}\hat{r}_\mathrm{c}^{-1},
\end{align}
\begin{equation}\label{eq:275} 
\sigma_3(\hat{r}_2)=3\int_0^{\hat{r}_2} d\hat{r}\,\hat{r}^2\rho(\hat{r})
=-\frac{\hat{T}}{2}\,\hat{r}^2\,\frac{\hat{z}'(\hat{r})}{\hat{z}(\hat{r})}.
\end{equation}
Expression (\ref{eq:274}), effectively a single integral, can be further
simplifed by eliminating $\hat{z}'(\hat{r})$ from (\ref{eq:275}) via an
integration by parts.
The result,
\begin{align}\label{eq:279}
\hat{W} =&-9g_s\hat{T}^{5/2}\int_0^{\hat{R}}d\hat{r}\,\hat{r}^2\,
g_{5/2}\big(\hat{z}(\hat{r})\big) \nonumber \\
&\hspace{10mm}+3g_s\hat{T}^{5/2}\hat{R}^3g_{5/2}\big(\hat{z}(\hat{R})\big)
+\frac{6}{5}\hat{r}_\mathrm{c}^{-1},
\end{align}
connects with the well-know expression infered from the viral theorem as explained in \cite{sgcfd} for the FD gas.

\subsubsection{Initial conditions}\label{sec:ini-con}
In all three cases, the composite macrostate at given temperature $\hat{T}$ is
specified by a one-parameter family of initial conditions for the gaseous halo,
namely by (\ref{eq:242}) for planar symmetry, (\ref{eq:262}) for cylindrical
symmetry, and (\ref{eq:282}) for spherical symmetry.
The parameter is  the interface fugacity $\hat{z}(\hat{r}_\mathrm{b})$.
In the temperature regime, where composite macrostates are realized, the one
representing thermal equilibrium has the lowest free energy.
Inspection shows that it is always associated with $\hat{z}(\hat{r}_\mathrm{b})=1$, implying that the gas is critical at the interface.

\subsection{BEC radius}\label{sec:BEC-rad}
All available evidence suggests that the density of the BEC is much higher than the density of the coexisting BE gas.
Any mixed-phase state in thermal and mechanical equilibrium thus consists of a BEC core and a gaseous halo.
Our analysis of macroscopic BE clusters employs a length scale, $r_\mathrm{s}$,
tailored to the description of gaseous density profiles, which is huge compared
to the natural length scale appropriate for the analysis of a pure BEC profile,
e.g. the scale $r_\mathrm{GP}$ inferred from a dimensional analysis of the
Gross-Pitaevski equation. 
We have found that $r_\mathrm{GP}/r_\mathrm{s}\sim N^{-\alpha}$ with $\alpha>0$ for all combinations of $\mathcal{D}$ and $\mathcal{D}_\sigma$.

A study of BEC density profiles in self-gravitating clusters of macroscopic size, which calculates density profiles on a much contracted length scale, requires the results of this work as an input, specifically the pressure at the interface produced by the weight of the gaseous halo, which depends on the gaseous density profile.

The present study, on the other hand, cannot ignore the spatial extension of the BEC altogether. 
It is a necessary and natural agent of short-distance regularization to prevent a divergent potential energy in $\mathcal{D}_\sigma\geq2$.
Our choice of reference state in Sec.~\ref{sec:free-ener} accommodates this need. 
By setting $0<\hat{r}_\mathrm{c}\ll1$ we simulate the presence of a provisional BEC with uniform density. 
The divergences avoided by this means are looming in the free energy expressions
(\ref{eq:257}) and (\ref{eq:274}) as well as in the boundary conditions
(\ref{eq:262}) and (\ref{eq:282}).
The impact of such regularizations will be further discussed case by case.


%
\section{Planar symmetry}\label{sec:planar}
%
The analysis of density profiles of BE clusters with planar symmetry starts from Eqs.~(\ref{eq:18})-(\ref{eq:19}) for ${\mathcal{D}_\sigma=1}$.
We set $g_s=1$ henceforth.
Gas clusters with planar symmetry are known to be stable against evaporation and also stable against gravitational collapse.
No wall confinement and short-distance regularization are needed. We set 
$\hat{r}_\mathrm{c}=0$ and (effectively) $\hat{R}=\infty$ throughout
Sec.~\ref{sec:planar}.

The exact density profile for planar MB clusters does not depend on $\mathcal{D}$: \cite{spitzer, camm, rybicki, kl, sc, cmct, selgra}.
\begin{equation}\label{eq:347} 
\rho(\hat{r})_\mathrm{MB}=\frac{1}{\hat{T}}\,\mathrm{sech}^2\!\left(\frac{\hat{r}}{\hat{T}}\right),
\end{equation}
With decreasing $\hat{T}$, it gradually becomes narrower and more strongly peaked at the central plane of the cluster.
Deviations of the BE density profiles in $\mathcal{D}=1,2,3$ emerge gradually, at first near the center of the cluster as a density enhancement. 
Whereas the central density of (\ref{eq:347}) diverges at $\hat{T}=0$, it does so for the BE clusters in $\mathcal{D}=1,2$ at a nonzero temperature, when condensation begins. 
In $\mathcal{D}=3$ the onset of condensation happens at finite central density.

The solution of Eqs.~(\ref{eq:18})-(\ref{eq:19}) for ${\mathcal{D}_\sigma=1}$ is reducible to quadrature when transcribed to an effectively $1^\mathrm{st}$-order ODE for the inverse function, $\hat{r}(\hat{\mu})$, of the scaled chemical potential $\hat{\mu}(\hat{r})=\hat{T}\ln\hat{z}(\hat{r})$.
That solution for a purely gaseous profile reads
\begin{subequations}\label{eq:365}
\begin{align}
& \hat{r}(\hat{\mu})=\int_{\hat{\mu}_0}^{\hat{\mu}}
d\hat{\mu}'\hat{s}(\hat{\mu}')\quad :~  \hat{\mu}\leq\hat{\mu}_0\leq0, \\
& \hat{s}({\hat{\mu})=-\frac{1}{\sqrt{2[a(\hat{\mu}_0)-a(\hat{\mu})]}}}, \\
& a(\hat{\mu})-a(\hat{\mu}_0)=
2\hat{T}^{\mathcal{D}/2}\int_{\hat{\mu}_0}^{\hat{\mu}}d\mu\,
g_{\mathcal{D}/2}\big(e^{\mu/\hat{T}}\big),
\end{align}
\end{subequations}
where the central chemical potential $\hat{\mu}_0$ is determined  from the integral,
\begin{equation}\label{eq:366}
\hat{T}^{\mathcal{D}/2}\int_{\hat{\mu}_0}^{-\infty} d\hat{\mu}\,g_{\mathcal{D}/2}\big(e^{\hat{\mu}/\hat{T}}\big)\hat{s}(\hat{\mu})=1.
\end{equation}

\subsection{$\mathcal{D}=1$}\label{sec:pla-D1}
\begin{figure}[t]
  \begin{center}
 \includegraphics[width=55mm]{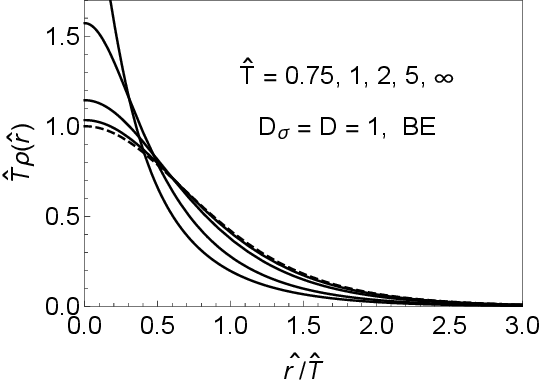}  
\end{center}
\caption{Rescaled density profiles of the BE gas in $\mathcal{D}_\sigma=\mathcal{D}=1$ at high $\hat{T}$. The dashed line represents the MB profile (\ref{eq:347}) at any $\hat{T}$ and the BE profile at $\hat{T}=\infty$.}
  \label{fig:1}
\end{figure}

At sufficiently high $\hat{T}$, the density profile has a smooth maximum at the center of the cluster and decays exponentially with distance $\hat{r}$.
We connect the BE profile with the MB profile (\ref{eq:347}) by plotting $\hat{T}\rho$ vs $\hat{r}/\hat{T}$ in Fig.~\ref{fig:1}.
The MB profile is invariant in this representation.
With $\hat{T}$ decreasing, the BE profile begins to deviate by an enhanced crowding at the center of the cluster.

\begin{figure}[b]
  \begin{center}
 \includegraphics[width=43mm]{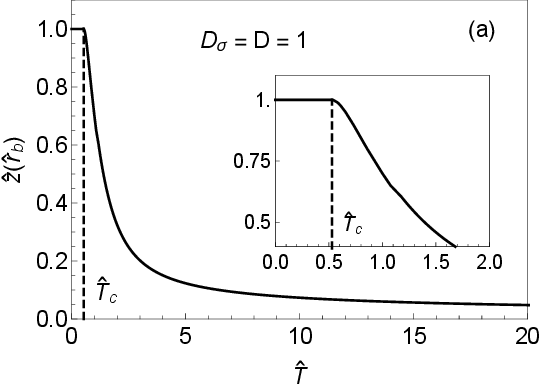} \hspace{0mm}%
  \includegraphics[width=42mm]{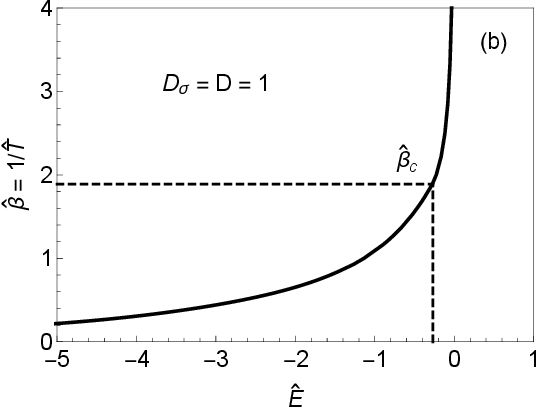}  
\includegraphics[width=42mm]{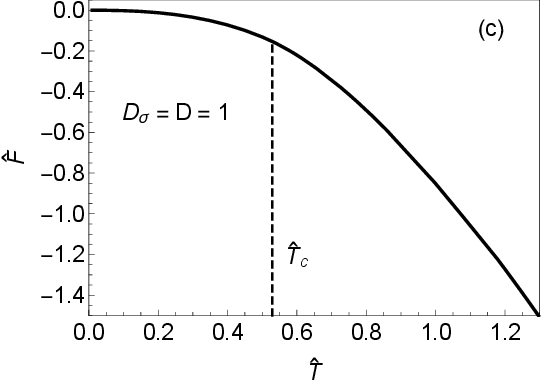} \hspace{0mm}%
  \includegraphics[width=42mm]{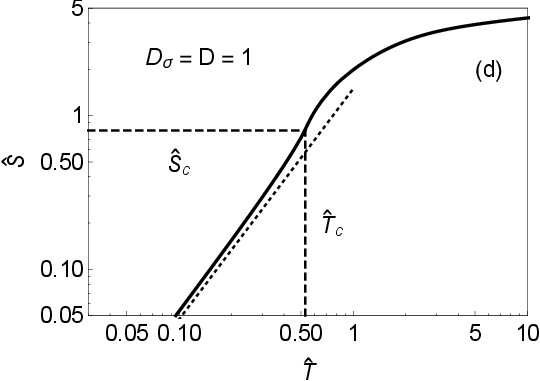}    
\end{center}
\caption{(a) Initial gaseous fugacity $\hat{z}(\hat{r}_\mathrm{b})$ vs $\hat{T}$
(here $\hat{r}_\mathrm{b}=0$). 
(b) Caloric curve: inverse temperature versus internal energy.  
(c) Helmholtz free energy versus temperature.  (d) Entropy versus temperature.
The dotted line highlights the asymptotic power law (\ref{eq:352}).}
  \label{fig:2}
\end{figure}

Our analysis produces a unique normalizable solution for $\hat{z}(\hat{r})$, representing a gaseous cluster over a range of temperatures. 
The central fugacity $\hat{z}(0)$  increases monotonically as $\hat{T}$ is lowered [Fig.~\ref{fig:2}(a)], reaching the critical value, $\hat{z}(0)=1$, at the temperature
\begin{equation}\label{eq:348} 
\hat{T}_\mathrm{c}\simeq 0.5287.
\end{equation}
The critical density profile diverges for $\hat{r}\to0$.
Condensation begins at the center of the cluster.
At $\hat{T}\leq\hat{T}_\mathrm{c}$ the interface fugacity is locked in to the critical value.
The mixed-phase macrostate with $\hat{z}(0)=1$ has the lowest
free energy.
It is unique for all subcritical temperatures.

The caloric curve [Fig.~\ref{fig:2}(b)]  has a discontinuity in slope at $\hat{T}_\mathrm{c}$.
The bottom portion represents a pure gas and the top portion a mixed-phase state.
It will be a useful benchmark for more complex caloric curves to be encountered later.
The free energy [Fig.~\ref{fig:2}(c)] is a monotonically decreasing function and has an imperceptibly weak singularity at $\hat{T}_\mathrm{c}$. 
The (negative) slope of that curve represents the variation of the entropy
with temperature.

\begin{figure}[b]
  \begin{center}
 \includegraphics[width=43mm]{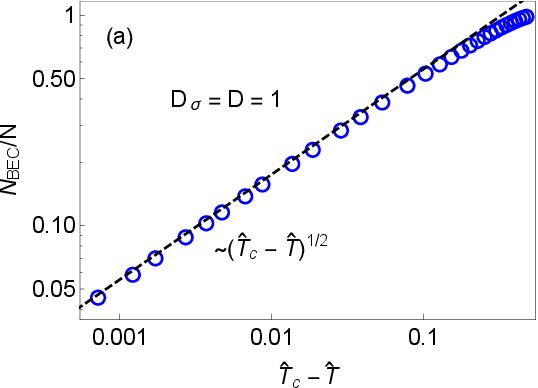} \hspace{0mm}%
  \includegraphics[width=42mm]{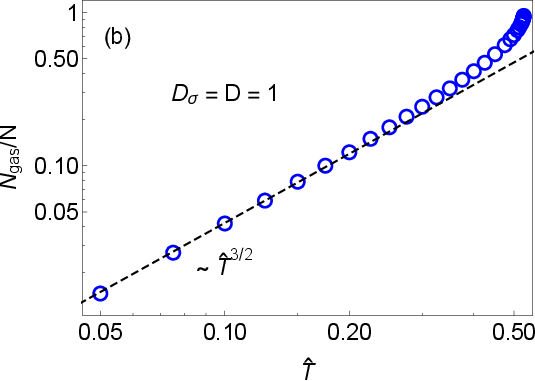}     
\end{center}
\caption{Power-law cusp singularities of (a) the order parameter $N_\mathrm{BEC}/N$ near $\hat{T}_\mathrm{c}$ and (b) its deviation from saturation, $N_\mathrm{gas}/N=1-N_\mathrm{BEC}/N$, near $\hat{T}=0$.}
  \label{fig:3plus}
\end{figure}

The entropy curve, plotted in Fig.~\ref{fig:2}(d) on doubly logarithmic scales, remains continuous.
It has a discontinuity in slope at $\hat{T}_\mathrm{c}$.
The subcritical entropy approaches zero as a power law, asymptotically for $\hat{T}\to0$:
\begin{equation}\label{eq:352} 
\hat{S} ~\sim~ \hat{T}^{\mathcal{D}/2+1}\quad :~
\hat{T}\ll\hat{T}_\mathrm{c}.
\end{equation}
This result turns out to be valid for all cases.
The order parameter $N_\mathrm{BEC}/N$  reaches saturation in a power-law cusp
with the same exponent [Fig.~\ref{fig:3plus}(b)].
At $\hat{T}_\mathrm{c}$ it reaches zero continuously in a square-root cusp [Fig.~\ref{fig:3plus}(a)]. 
The exponent of this singularity does not change with $\mathcal{D}$.

The onset of condensation has the hallmarks of a second-order phase transition.
For the investigation of additional critical singularities, we start from the ODE for the chemical potential inferred from (\ref{eq:18}),
\begin{subequations}\label{eq:354}
\begin{align}\label{eq:354a}
& \hat{\mu}''+2\hat{T}^{\mathcal{D}/2}g_{\mathcal{D}/2}\big(e^{\hat{\mu}/\hat{T}}\big)=0,
\\ \label{eq:354b}
& \hat{\mu}(0)\geq0,\quad \hat{\mu}'(0)=0.
\end{align}
\end{subequations}
Expanding the BE function (\ref{eq:3}) near criticality, $\hat{\mu}\to0$, leads (for $\mathcal{D}=1$)  to the simplified (and rescaled) ODE,
\begin{equation}\label{eq:356} 
\bar{\mu}_\mathrm{s}''=\bar{\mu}_\mathrm{s}^{-1/2},\quad 
\bar{\mu}_\mathrm{s}\doteq-\big(2\hat{T}_\mathrm{c}\sqrt{\pi}\big)^{-2/3}\hat{\mu}_\mathrm{s},
\end{equation}
with initial conditions $\bar{\mu}_\mathrm{s}(0)=0$, $\bar{\mu}_\mathrm{s}'(0)\geq0$.
The exact solution,
\begin{equation}\label{eq:357}
\bar{\mu}_\mathrm{s}(\hat{r})=\left(\frac{3\hat{r}}{2}\right)^{4/3},
\end{equation}
represents the leading singularity in the form of a power-law cusp.
The critical divergence in the density profile then follows immediately:
\begin{equation}\label{eq:358}
\rho_\mathrm{s}(\hat{r}) \sim \hat{r}^{-2/3}.
\end{equation}
The cusp singularity of (\ref{eq:357}) is sufficiently weak to make $\bar{\mu}'(0)=0$, which is consistent with $\hat{z}'(0)=0$ as invoked earlier for the critical fugacity.

Only at $\hat{R}\lesssim1$ does the value of $\hat{T}_\mathrm{c}$ acquire a significant $\hat{R}$-dependence. 
Furthermore, the numerical evidence suggests that $\hat{T}_\mathrm{c}$ diverges under very tight confinement, $\hat{R}\simeq5\times10^{-4}$.

\subsection{$\mathcal{D}=2$}\label{sec:pla-D2}
Increasing the dimensionality $\mathcal{D}$ while maintaining the planar symmetry ($\mathcal{D}_\sigma=1$) produces, for the most part, systematic quantitative changes. 
The approach of the BE density profiles toward the universal MB profile (\ref{eq:347}) at high $\hat{T}$ is qualitatively similar to, but faster than in the case $\mathcal{D}=1$.
Criticality is reached at a higher temperature,
\begin{equation}\label{eq:359} 
\hat{T}_\mathrm{c}\simeq 0.7797.
\end{equation}
The low-temperature asymptotics of the order parameter and the entropy are again governed by (\ref{eq:352}).
The plots for the central fugacity, the caloric curve, the free energy, and the entropy look very similar to the results shown in Fig.~\ref{fig:2}.
The discontinuity in slope at $\hat{T}_\mathrm{c}$ in the caloric curve is more pronounced.

Qualitative changes make their appearance in critical singularities.
For the case $\mathcal{D}=2$, we extract from Eqs. (\ref{eq:354}) the ODE,
\begin{equation}\label{eq:376}
\bar{\mu}_\mathrm{s}''+2\ln\bar{\mu}_\mathrm{s}=0,\quad 
\bar{\mu}_\mathrm{s}\doteq-\frac{\hat{\mu}_\mathrm{s}}{\hat{T}_\mathrm{c}}
\end{equation}
for the leading singularity of the chemical potential.
The exact solution in this case, 
\begin{equation}\label{eq:377}
\bar{\mu}_\mathrm{s}(\hat{r})=
\exp\left(1-2\left[\mathrm{erf}^{-1}\left(1-\frac{\hat{r}}{\sqrt{e\pi/2}}\right)\right]^2\right),
\end{equation}
encodes a more complex cusp with logarithmic corrections.
The leading singularity in the density profile thus turns out to be a logarithmic divergence:
\begin{equation}\label{eq:378}
\rho(\hat{r})\simeq -\hat{T}_\mathrm{c}\ln\bar{\mu}_\mathrm{s}\sim \sqrt{|\ln\hat{r}|}
+\mathrm{O}\big(\sqrt{\ln|\ln\hat{r}|}\big).
\end{equation}
The implicit exact solution (\ref{eq:365}) for density profiles can, in $\mathcal{D}=2$, be made more explicit in parametric form, 
\begin{subequations}\label{eq:381}
\begin{align}
& \rho(\bar{\mu})=-\hat{T}\ln\big(1-e^{-\bar{\mu}}\big), \\
& \hat{r}(\bar{\mu})=\int_{\bar{\mu}_0}^{\bar{\mu}}\frac{d\mu}{2\sqrt{a(\bar{\mu}_0)-a(\mu)}}, \\
& a(\bar{\mu})\doteq \frac{1}{2}\bar{\mu}^2
+\bar{\mu}\ln\big(1-e^{-\bar{\mu}}\big)-b(\bar{\mu}), \\
& b(\bar{\mu})\doteq 
\ln\big(1-e^{\bar{\mu}}\big)+\mathrm{Li}_2(e^{\bar{\mu}}),\quad 
\bar{\mu}\doteq-\hat{\mu}/\hat{T},
\end{align}
\end{subequations}
where the dependence of $\bar{\mu}_0$ on $\hat{T}\geq\hat{T}_\mathrm{c}$ is
determined by the normalization condition (\ref{eq:366}), here rendered as,
\begin{equation}\label{eq:382}
\int_{\bar{\mu}_0}^\infty d\mu\,\frac{\ln\big(1-e^{-\mu}\big)}
{2\sqrt{a(\bar{\mu}_0)-a(\mu)}}=\frac{1}{\hat{T}}.
\end{equation}
The logarithmic terms are characteristic for $\mathcal{D}=2$.
We shall encounter them again in $\mathcal{D}=2$ for clusters with cylindrical symmetry ($\mathcal{D}_\sigma=2$).
 
\subsection{$\mathcal{D}=3$}\label{sec:pla-D3}
The trend noted in Sec. \ref{sec:pla-D2} continues as we add another spatial
dimension and keep the planar symmetry of the cluster. 
The approach to the MB profile (\ref{eq:347}) at high $\hat{T}$ is yet faster as is the approach to criticality when $\hat{T}$ is lowered.
Condensation sets in earlier, at 
\begin{equation}\label{eq:383} 
\hat{T}_\mathrm{c}\simeq 0.88913.
\end{equation}
The curves such as shown in Figs.~\ref{fig:1} and \ref{fig:2} for $\mathcal{D}=1$ are again similar for $\mathcal{D}=3$, but with a yet more pronounced kink in the caloric curve.
A distinctive feature of the case $\mathcal{D}=3$ for clusters of any symmetry is that the critical gas density is finite (non-divergent).
An inspection of the ODE (\ref{eq:354}) for $\mathcal{D}_\sigma=1$ and $\mathcal{D}=3$ reveals (see Appendix~\ref{sec:appa}) that the critical chemical potential can be expanded into a power series beginning with the quadratic term:
\begin{align}\label{eq:419}
& \bar{\mu}(\hat{r})\doteq -\frac{\hat{\mu}}{\hat{T}_\mathrm{c}}
=\sum_{n=2}^\infty a_n\hat{r}^n, \nonumber \\
& a_2(\hat{T}_\mathrm{c})=\hat{T}_\mathrm{c}^{1/2}\, \zeta \left(\textstyle \frac{3}{2}\right), 
\quad a_3(\hat{T}_\mathrm{c})=-\textstyle \frac{2}{3}\,\hat{T}_\mathrm{c}^{3/4} 
\sqrt{\pi  \zeta \left(\textstyle \frac{3}{2}\right)}, \nonumber \\
& a_4(\hat{T}_\mathrm{c})={\textstyle \frac{1}{18} \hat{T}_\mathrm{c} \Big[2\pi -3\zeta \left(\textstyle\frac{1}{2}\right) \zeta
   \left(\textstyle\frac{3}{2}\right)\Big]}, \nonumber \\
& a_5(\hat{T}_\mathrm{c})= {\textstyle \frac{1}{12}} \hat{T}_\mathrm{c}^{5/4} \zeta \left(\textstyle \frac{1}{2}\right)
   \sqrt{\pi  \zeta \left(\textstyle\frac{3}{2}\right)}, \quad ...
\end{align}
A series beginning with zeroth power follows for the critical density:
\begin{align}\label{eq:420}
& \rho(\hat{r}) = \sum_{n=0}^\infty c_n\hat{r}^n, \quad
c_0=\hat{T}_\mathrm{c}^{3/2} \zeta \left(\textstyle\frac{3}{2}\right), \nonumber \\
& c_1=-2 \hat{T}_\mathrm{c}^{7/4} \sqrt{\pi  
\zeta \left(\textstyle\frac{3}{2}\right)}, \nonumber \\
& c_2 ={\textstyle\frac{1}{3}}\hat{T}_\mathrm{c}^2 
   \left[2\pi -3\zeta \left(\textstyle\frac{1}{2}\right) \zeta
   \left(\textstyle\frac{3}{2}\right)\right],\quad \ldots
\end{align}
The critical density in $\mathcal{D}=3$ has linear cusp at $\hat{r}=0$.

\subsection{Salient features}\label{sec:sal-fea-plan}
MB particles are (effectively) point particles, whereas BE particles are not.
The differences in statistics manifest themselves when the local density is
sufficiently high to make the mean interparticle distance comparable 
to the thermal wavelength. 
The distinctive attribute of BE statistics is the multiple occupancy of one-particle levels combined with the indistinguishability of many-particle states with identical occupancies.

The universal shape of the MB density profile (\ref{eq:347}) with a smooth central maximum and exponential tails is shared by BE clusters at high $\hat{T}$.  
Upon lowering $\hat{T}$, the MB density profile smoothly grows in height and shrinks in width, approaching a $\delta$-function as $\hat{T}\to0$.
The BE density profile shows a similar trend initially, but with an enhanced particle concentration near the center of the cluster.
Unlike the MB profile, the BE profile acquires, at a nonzero  $\hat{T}_\mathrm{c}$, a singularity at the center of the cluster, where the density is highest.

Whereas the dimensionality $\mathcal{D}$ of the space has no impact on the shape of the MB profile for planar clusters (or clusters of any other symmetry), that is not the case for BE profiles.
As $\hat{T}$ approaches $\hat{T}_\mathrm{c}$ from above, the central density has a power-law divergence, $\sim\hat{r}^{-2/3}$, in $\mathcal{D}=1$, a logarithmic divergence, $\sim\sqrt{|\ln\hat{r}|}$, in $\mathcal{D}=2$, and a linear cusp singularity, $\sim a-b\hat{r}$, in $\mathcal{D}=3$.
The value of $\hat{T}_\mathrm{c}$ increases with $\mathcal{D}$.
The variation of critical singularities with $\mathcal{D}$ may be unusual for a mean-field context, but the trends are in line with the expectation that the strength of fluctuations are strongest for the lowest $\mathcal{D}$.

The singularities of the BE gas at $\hat{T}_\mathrm{c}$, which marks the onset of condensation, bear the hallmarks of a second-order phase transition in dimensions $\mathcal{D}=1,2,3$.
The order parameter, represented by the fraction of particles in the ground state, rises continuously from zero in a cusp singularity. 
The entropy has a discontinuity in slope.
Condensation is associated with a drastic change in length scale, which is only summarily accounted for in this study as explained earlier (Sec.~\ref{sec:BEC-rad}).
The caloric curves of MB and BE clusters are both monotonic across the full temperature range, thus ruling out any form of mechanical instability (gravitational collapse). 
Only the BE caloric curve has a (cusp) singularity.

%
\section{Cylindrical symmetry}\label{sec:cylind}
%
The precarious stability of gas clusters with $\mathcal{D}_\sigma=2$ against collapse  or evaporation is well established. 
We shall work with a nonzero reference radius $\hat{r}_\mathrm{c}$ to manage singularities associated with the former and a finite radius of confinement $\hat{R}$ against the latter.

The MB gas is again a useful benchmark.
A cylindrical MB cluster is stable against collapse above the ($\hat{R}$-independent) threshold
temperature \cite{cf, stodolkiewicz, ostriker, salzberg, klb, paddy2d, aly, ar, ap, sc, cmct, bppv, virialD, selgra}
\begin{equation}\label{eq:394}
\hat{T}_\mathrm{MB}=\frac{1}{2}.
\end{equation}
The exact density inferred from (\ref{eq:21}) 
for $\hat{T}>\hat{T}_\mathrm{MB}$ is \cite{sc,selgra}:
\begin{equation}\label{eq:395} 
 \rho(\hat{r})_\mathrm{MB}=\frac{1}{\hat{R}^2}\frac{4\hat{T}\big(\hat{T}-\hat{T}_\mathrm{MB}\big)}
 {\big[(\hat{r}/\hat{R})^2+2(\hat{T}-\hat{T}_\mathrm{MB})\big]^2}.
\end{equation}
This profile becomes sharply peaked at $\hat{r}=0$ when $\hat{T}$  (at fixed $\hat{R}$) approaches $\hat{T}_\mathrm{MB}$ from above.
The limits ${\hat{R}\to\infty}$ and $\hat{T}\to\hat{T}_\mathrm{MB}$ are not interchangeable.
Taking the combined limit,
\begin{equation}\label{eq:431} 
\hat{R}\to\infty,\quad \hat{T}\to\hat{T}_\mathrm{MB},\quad
\frac{\hat{T}^2}{4\hat{R}^2(\hat{T}-\hat{T}_\mathrm{MB})}=c>0,
\end{equation}
in expression (\ref{eq:395}), produces a one-parameter family of critical MB profiles \cite{cmct,selgra}:
\begin{equation}\label{eq:432} 
\rho_\mathrm{c}(\hat{r})_\mathrm{MB}=\frac{4c}{\hat{T}_\mathrm{MB}}
\left[1+2c\left(\frac{\hat{r}}{\hat{T}_\mathrm{MB}}\right)^2\right]^{-2}.
\end{equation}
The central density, $\rho_\mathrm{c}(0)_\mathrm{MB}=4c/\hat{T}_\mathrm{MB}$, can assume any non-negative value.

The BE gas is known to exert a lower pressure than the MB gas does under equivalent circumstances.
Upon cooling, it gives in to gravity earlier and differently.
The dimensionality $(\mathcal{D}=2,3)$ of the space in which cylindrical clusters $(\mathcal{D}_\sigma=2)$ are realized  matters.
For the sake of brevity, we focus on the case $\mathcal{D}=2$.
The analysis again starts from Eqs.~(\ref{eq:18})-(\ref{eq:19}).

\subsection{Onset of condensation}\label{sec:cyl-onset}
\begin{figure}[htb]
  \begin{center}
 \includegraphics[width=55mm]{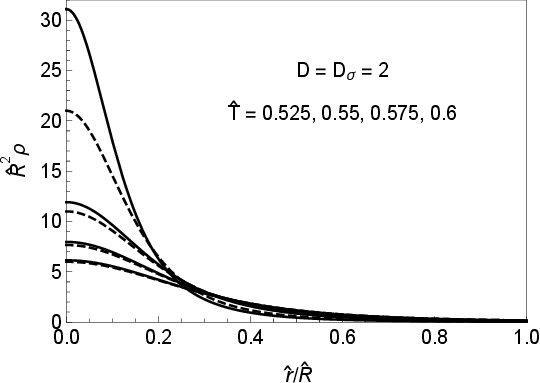}
\end{center}
\caption{Rescaled density profile of the BE gas in $\mathcal{D}=\mathcal{D}_\sigma=2$ confined to radius $\hat{R}=20$. The dashed lines represent the MB profile (\ref{eq:395}), which are independent of $\hat{R}$ in this plot.}
  \label{fig:3}
\end{figure}

The emergent deviations of the BE density profile from the MB benchmark profile (\ref{eq:395}) in the high-temperature regime are illustrated in Fig.~\ref{fig:3}.
The scales used render the MB profiles independent of the radius of
confinement $\hat{R}$.
A large $\hat{R}$ for the BE gas, makes the emerging differences conspicuous near $\hat{T}_\mathrm{MB}$.
The higher compressibility of the BE gas is manifest in the enhanced density near the center of the cluster, an attribute already observed in planar clusters.

\begin{figure}[htb]
  \begin{center}
 \includegraphics[width=41mm]{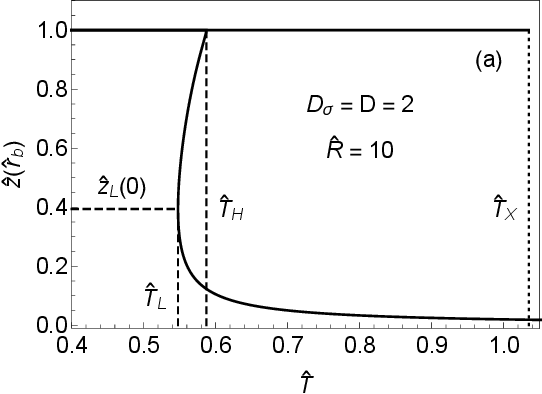}
 \includegraphics[width=43mm]{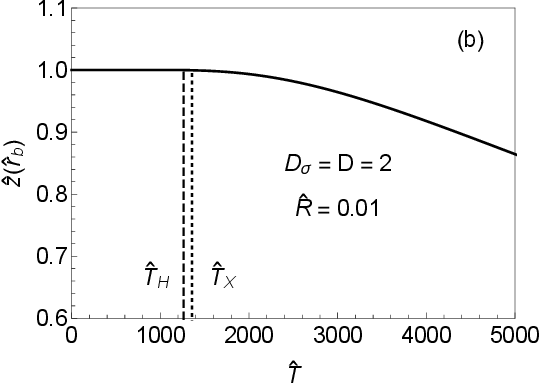}
\end{center}
\caption{(a) Initial fugacity $\hat{z}(\hat{r}_\mathrm{b})$ of the gas part in a cylindrical BE cluster for $\mathcal{D}=\mathcal{D}_\sigma=2$ confined to radius (a) $\hat{R}=10$ and (b) $\hat{R}=0.01$.
The curves represent a pure gas. The horizontal segments at $\hat{z}(\hat{r}_\mathrm{b})=1$ represent solutions consisting of a BEC surrounded by a gasous halo. 
The reference states in use have radius (a) $\hat{r}_\mathrm{c}=10^{-1}$ and (b) $\hat{r}_\mathrm{c}=10^{-4}$, equal in ratio as the radius of confinement $\hat{R}$.}
  \label{fig:4}
\end{figure}

At lower temperatures, the evolution of cylindrical BE profiles is rich and varies with the radius of confinement.
There are two regimes, in our analysis represented by the cases
$\hat{R}=10$ and $\hat{R}=0.01$.
The curves in Fig.~\ref{fig:4} show the fugacity $\hat{z}(\hat{r}_\mathrm{b})$ at the center of the cluster if it is purely gaseous $(\hat{r}_\mathrm{b}=0)$ or at the interface between the BEC and the gaseous halo $(\hat{r}_\mathrm{b}\neq 0)$.
In both regimes, a two-phase solution with $\hat{z}(\hat{r}_\mathrm{b})=1$ exists for $0<\hat{T}<\hat{T}_\mathrm{X}$ with the central fugacity reaching criticality at $\hat{T}_\mathrm{H}$.

Only in the first regime do distinct solutions of pure gas exist.
Here we note four temperature intervals, delimited by $\hat{T}_\mathrm{L}\simeq 0.547$, $\hat{T}_\mathrm{H}\simeq 0.587$, and $\hat{T}_\mathrm{X}\simeq1.035$.
All three are higher than $\hat{T}_\mathrm{MB}$, but the lower two not by much.
A unique gaseous solution exists at $\hat{T}>\hat{T}_\mathrm{X}$ and a unique two-phase solution with a critical interface at $\hat{T}<\hat{T}_\mathrm{L}$.
At intermediate temperatures, three solutions coexist, a pair of gaseous solutions and one two-phase solution for $\hat{T}_\mathrm{L}<\hat{T}<\hat{T}_\mathrm{H}$, or a pair of two-phase solutions and one gaseous solution for $\hat{T}_\mathrm{H}<\hat{T}<\hat{T}_\mathrm{X}$.

When we gradually shrink the confining radius $\hat{R}$, the values of $\hat{T}_\mathrm{L}$, $\hat{T}_\mathrm{H}$, and $\hat{T}_\mathrm{X}$ both increase at different rates.  
The first two merge into $\hat{T}_\mathrm{H}$ at 
\begin{equation}\label{eq:400} 
\hat{R}_1\simeq 0.053,
\end{equation}
which marks the border to the second regime, characterized by tight confinement.
Here the central fugacity becomes critical, $\hat{z}_\mathrm{L}(0)\to1$.

In the second regime, for $\hat{R}=0.01$, there are only three temperature intervals, delimited by $ \hat{T}_\mathrm{H}\simeq 1263$ and $\hat{T}_\mathrm{X}\simeq 1357$. 
A unique gaseous (mixed) solution exists at $\hat{T}>\hat{T}_\mathrm{X}$ $(\hat{T}<\hat{T}_\mathrm{H})$ and a coexistence of two mixed and one gaseous solution at $\hat{T}_\mathrm{H}<\hat{T}<\hat{T}_\mathrm{X}$.

\subsection{Mechanical instabilities}\label{sec:cyl-mech-inst}
Next we take a closer look at the mechanical stability and thermal equilibrium of coexisting macrostates in the interval $\hat{T}_\mathrm{L}<\hat{T}<\hat{T}_\mathrm{X}$ for the case $\hat{R}=10$ of the first regime. 
What happens in the second regime is a mere simplification on account of the merger $\hat{T}_\mathrm{L}\to\hat{T}_\mathrm{H}$.

For the examination of stability conditions we use the free energy expressions
from Sec.~\ref{sec:Ds=2} and also employ caloric curves.
Unlike in systems of planar symmetry, we must set $\hat{r}_\mathrm{c}>0$ in order to avoid a divergent potential energy and to avoid a divergent boundary condition (\ref{eq:262}). 

\begin{figure}[t]
  \begin{center}
 \includegraphics[width=42mm]{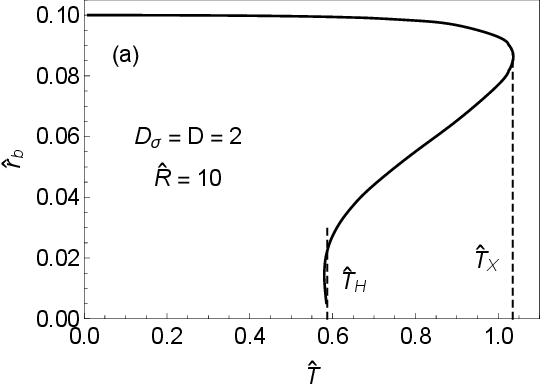}%
 \includegraphics[width=42mm]{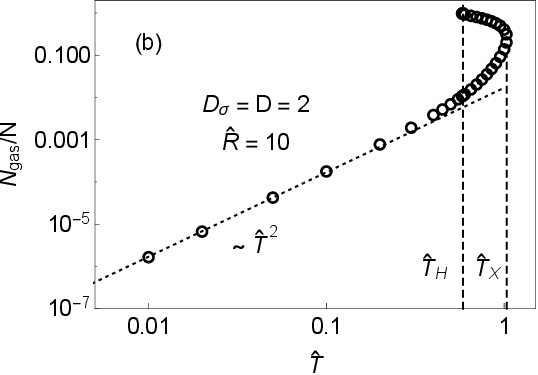}
 \end{center}
\caption{(a) Dependence of interface radius $\hat{r}_\mathrm{b}$ on temperature
$\hat{T}$. (b) Fraction of particles in the gas phase as a function of $\hat{T}$
for all identified mixed-state solutions. The dashed line indicates
$\sim\hat{T}^2$ asymptotics. The reference radius has been set to
$\hat{r}_\mathrm{c}=0.1$.}
  \label{fig:5}
\end{figure}

The two coexisting gaseous macrostates for $\hat{T}_\mathrm{L}<\hat{T}<\hat{T}_\mathrm{H}$ are readily identified in Fig.~\ref{fig:4}(a).
For the identification of the two coexisting mixed states for $\hat{T}_\mathrm{H}<\hat{T}<\hat{T}_\mathrm{X}$, we plot $\hat{r}_\mathrm{b}$ versus $\hat{T}$ in Fig.~\ref{fig:5}(a).
The two mixed macrostates are associated with the different values of the order parameter as shown in Fig.~\ref{fig:5}(b). 
There is even the hint of a third solution.
The upper branch of panel (a) corresponds to the lower branch in panel (b). 
Its extension toward zero temperature shows the quadratic low-$\hat{T}$ asymptotics.

\begin{figure}[htb]
  \begin{center}
 \includegraphics[width=43mm]{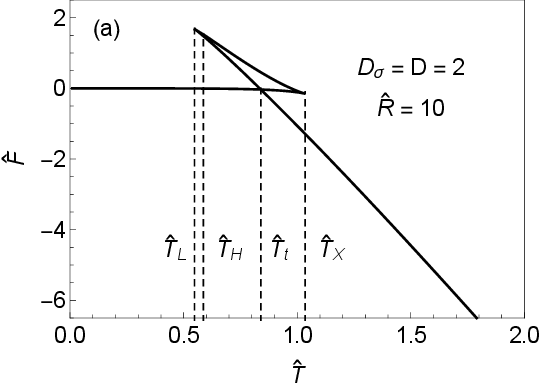}%
 \includegraphics[width=41mm]{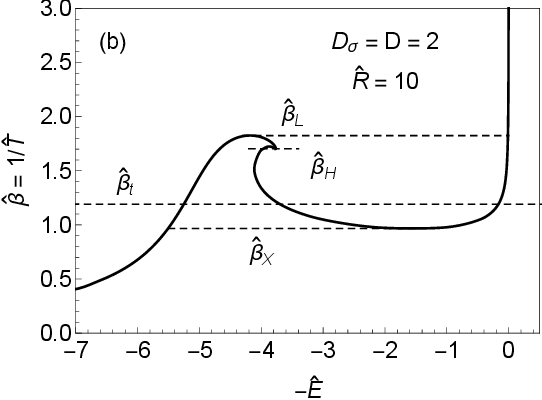}
 \end{center}
\caption{(a) Free energy versus temperature. (b) Caloric curve. Both sets include data for gaseous macrostates and data for the mixed macrostates (with $\hat{r}_\mathrm{c}=0.1$).}
  \label{fig:6}
\end{figure}

Among the coexisting macrostates for
$\hat{T}_\mathrm{L}<\hat{T}<\hat{T}_\mathrm{X}$, the ones representing
thermodynamic equilibrium are readily identified 
in the free energy plot of Fig.~\ref{fig:6}(a).
The lowest branch switches from a gaseous state to a mixed-phase state at $\hat{T}_\mathrm{t}\simeq0.84$.
In a homogeneous system, a first-order phase transition would be expected to take place at this temperature.
That is not the case here, as the caloric curve shown in Fig.~\ref{fig:6}(b) makes quite clear.

Self-gravitating clusters are inhomogeneous and have nontrivial mechanical stability conditions. 
The Poincar\'e criterion for stability limits in the caloric curve are points of zero slope (for systems analyzed as canonical ensembles). 
Such points exist for $\hat{T}_\mathrm{L}$, $\hat{T}_\mathrm{H}$, and $\hat{T}_\mathrm{X}$, but not for $\hat{T}_\mathrm{t}$.
Upon cooling from high $\hat{T}$, the gas phase is mechanically stable down to $\hat{T}_\mathrm{L}$, at which point it suffers an instability and settles in a mixed macrostate displaced horizontally to a lower energy on the right.
Conversely, when heat is added quasistatically from low $\hat{T}$, the mixed state loses mechanical stability at $\hat{T}_\mathrm{X}$ and settles, horizontally to the left, in a gaseous state of higher energy.

The temperature $\hat{T}_\mathrm{H}$, marked by a local minimum of the caloric
curve, comes into play as a point of mechanical instability only if a
macrostate nearby happens to be realized in some way.
The two thermodynamic equilibrium macrostates at the crossing point of the
free energy curve correspond to the outer intersection points of the dashed line
at
$\hat{\beta}_\mathrm{t}$ with the caloric curve.
They are without significance regarding mechanical stability.
Unlike in homogeneous systems, there is no quasistatic process at $\hat{T}_\mathrm{t}$ that connects the two equilibrium states.

\begin{figure}[htb]
  \begin{center}
 \includegraphics[width=42mm]{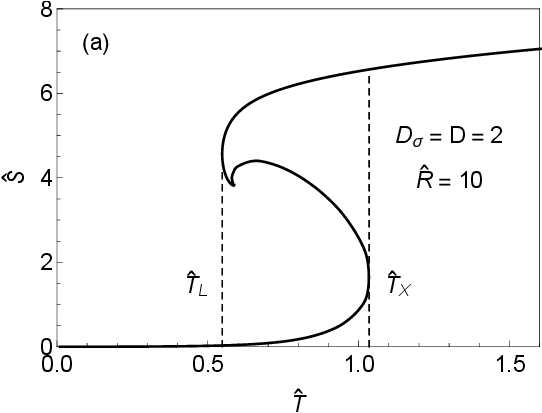}%
 \includegraphics[width=42mm]{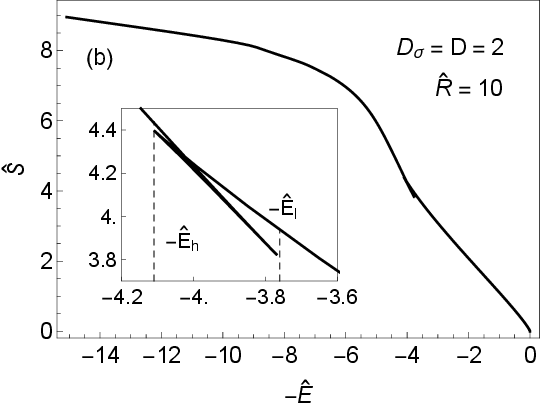}
 \end{center}
\caption{Entropy versus (b) temperature and (b) versus internal energy.}
  \label{fig:7}
\end{figure}

The mechanical instability at $\hat{T}_\mathrm{L}$ on the way down in temperature triggers processes that eject heat whereas the mechanical instability at $\hat{T}_\mathrm{X}$ on the way up triggers processes that absorb heat.
This is illustrated by the two dashed lines in the entropy versus temperature
plot of Fig.~\ref{fig:7}(a). 
Each instability begins at the tangent point and ends at the intersection point. 

Processes triggered by mechanical instabilities are likely to be fast and take their course at constant internal energy with little heat exchange between the system and the outside through the confining wall.
Therefore, it makes sense to describe the transitions between the gaseous state and the mixed-phase state in the microcanical ensemble.
For that purpose, we take a look at entropy plotted versus internal energy [Fig.~\ref{fig:7}(b)].

\begin{figure}[htb]
  \begin{center}
 \includegraphics[width=42mm]{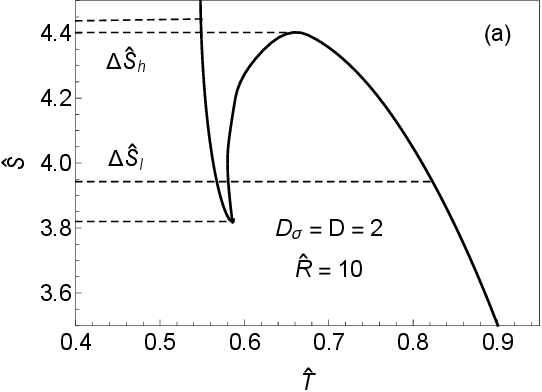}%
 \includegraphics[width=42mm]{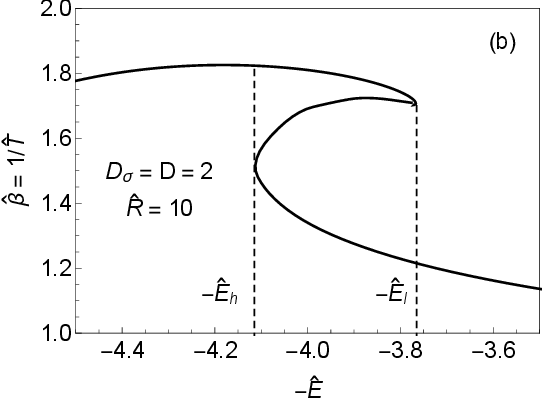}
 \end{center}
\caption{Partial views of (a) entropy versus temperature and (b) caloric curve.
Data previously 
used in Fig.~\ref{fig:7}(a) and \ref{fig:6}(b) are reproduced within zoomed-in
window.}
  \label{fig:8}
\end{figure}

The curve is largely monotonic except for a folded stretched with two hairpin turns at $E_\mathrm{h}$ and $E_\mathrm{l}$ as emphasized in the inset.
These values are locations of mechanical instabilities with no heat exchange. 
The process starts at the hairpin and ends where the dashed line intersects the curve.
Both processes are associated with an entropy increase as expected and are
identified in the zoomed-in entropy versus temperature plot of
Fig.~\ref{fig:8}(a). 
Each process begins at the tangent point and ends at the intersection point of a dashed line with the curve.
The entropy increases in both processes, but the temperature goes up in one and down in the other.

For systems analyzed as microcanonical ensembles, the Poincar\'e criterion for stability limits in the caloric curve are points of infinite slopes such as those identified in a zoomed-in version of the caloric curve shown in Fig.~\ref{fig:8}(b). 
Again the mechanical instabilities begin at the tangent points and end at the intersection points.

In summary, cylindrical BE clusters are precipitated into and out of condensation by way of mechanical instabilities with hysteretic features involved. 
In the canonical ensemble all the action happens between temperatures $\hat{T}_\mathrm{L}$ and $\hat{T}_\mathrm{X}$, in the microcanonical ensemble between energies $E_\mathrm{l}$ and $E_\mathrm{h}$. 
The latter are closer together on the caloric curve than the former.

\subsection{Limit of no confinement}\label{sec:lim-no-com}
A gradual widening of the radius of confinement, establishes a point of contact between the BE and MB phase behaviors.
The numerical evidence suggests the limits,
\begin{subequations}\label{eq:397}
\begin{align}\label{eq:397a} 
& \lim_{\hat{R}\to\infty}\hat{T}_\mathrm{L}=\hat{T}_\mathrm{MB}=\frac{1}{2}, \\
& \lim_{\hat{R}\to\infty}z_\mathrm{L}(0)=0, \\
& \lim_{\hat{R}\to\infty}\hat{T}_\mathrm{H}\doteq\hat{T}_\mathrm{H}^{(\infty)}=0.5629...,
\\
& \lim_{\hat{R}\to\infty}\hat{T}_\mathrm{X}\doteq\hat{T}_\mathrm{X}^{(\infty)}(\hat{r}_\mathrm{c})>\hat{T}_\mathrm{H}^{(\infty)}.
\end{align}
\end{subequations}
The gaseous macrostate at $\hat{T}>\hat{T}_\mathrm{L}$ has very low density, which makes it MB-like. 
It continues to exist down to near $\hat{T}_\mathrm{MB}$, where the MB cluster suffers a collapse. 
However, for the BE gas an alternative macrostate becomes available already at the higher temperature $\hat{T}_\mathrm{X}^{(\infty)}$, consisting of a BEC core surrounded by a gaseous halo.
The MB limit of the BE cluster with cylindrical symmetry is restricted in scope.
The limit $\hat{R}\to\infty$ of the BE cluster involves two subtleties.

(i) At $\hat{T}_\mathrm{L}$, where the gaseous solution disappears, the density at the center of the BE cluster is finite and approaching zero as $\hat{R}\to\infty$.
In the MB gas cluster, by contrast, the central density diverges when $\hat{T}$ approaches $\hat{T}_\mathrm{MB}$ for fixed $\hat{R}$.
This apparent contradiction is resolved by the one-parameter family of density profiles (\ref{eq:432}) for the MB cluster at $\hat{T}=\hat{T}_\mathrm{MB}$ and $\hat{R}=\infty$ with a range of central densities between zero and infinity.
The limit $\hat{R}\to\infty$ taken for the BE gas realizes one (extreme) value of this continuum.

(ii) The landmark temperature $\hat{T}_\mathrm{X}$ and its no-confinement limit $\hat{T}_\mathrm{X}^{(\infty)}$ are both dependent on the existence of a BEC with nonzero extension, in this work provisionally represented by the radius $\hat{r}_\mathrm{c}$ of the reference state.
Condensates have, of course, no part in an MB gas.
Estimates for $\hat{r}_\mathrm{c}$ in cylindrical BE clusters including the no-confinement limit have to await a theory of the self-gravitating BEC under the weight of a gaseous halo.
As mentioned earlier, such halos are analyzed in this work with provisional BECs.
Their specifications will be needed as input in a separate study dedicated to the density profile of BECs on a different length scale. 

\begin{figure}[htb]
  \begin{center}
  \includegraphics[width=55mm]{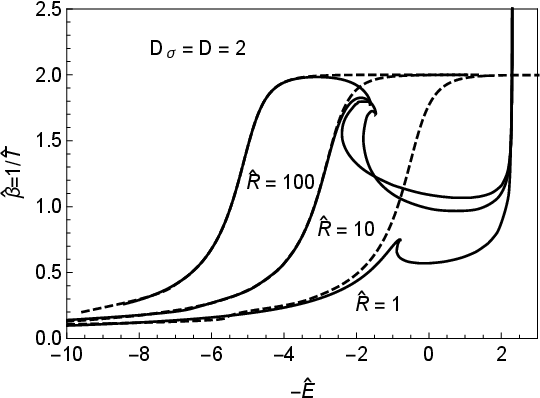}
 \end{center}
\caption{Caloric curves for clusters with different radii $\hat{R}$ of confinement in comparison. BE clusters are represented by solid lines and MB clusters by dashed lines. We use $\hat{r}_\mathrm{c}=0.1$ in all cases.}
  \label{fig:9}
\end{figure}

The limited scope of the MB limit in BE clusters is also evident in the
comparative plot of caloric curves presented in Fig.~\ref{fig:9}.
It highlights the impact of bosonic quantum statistics in clusters with cylindrical symmetry.
The MB curves are monotonically rising and leveling off at $\hat{T}_\mathrm{MB}$, signaling a gravitational collapse.
The internal energy $\hat{E}$ decreases gradually as $\hat{T}$ approaches that threshold.
At given $\hat{T}>\hat{T}_\mathrm{MB}$, $\hat{E}$ is lower if the wall confinement is tighter.
The threshold $\hat{T}_\mathrm{MB}$, on the other hand, does not depend on $\hat{R}$ because the gravitational collapse is counteracted by kinetic energy alone, which only depends on temperature. 

The BE gas mimics the MB gas for as long as the thermal wavelength is much shorter than the mean inter-particle distance. 
This is the case at high $\hat{T}$, where the curves overlap.
The BE caloric curve is not monotonic even for very tight confinement (not shown).
It features two sets of landmarks discussed earlier: (i) a smooth local maximum
and a smooth local minimum at finite internal energies, marking instabilities in
the framework of the canonical ensemble, (ii) two points of infinite slope,
positioned closer together, signaling mechanical 
instabilities in the framework of the microcanonical ensemble. 
The steeply rising portion of all BE caloric curves on the right represent the BEC, taken to have a high but finite density.

\subsection{Criticality}\label{sec:cyl-crit}
All phenomena described thus far are similar for clusters with $\mathcal{D}_\sigma=2$ in $\mathcal{D}=2,3$. 
The critical singularities, on the other hand, strongly depend on $\mathcal{D}$.
For all three cases with planar symmetry $(\mathcal{D}_\sigma=1)$, we were able to find an exact solution (Sec.~\ref{sec:planar}).
For all three cases in $\mathcal{D}=3$, the critical density profile  can be
expanded into a power series  (Appendix~\ref{sec:appa}).
That leaves the case $\mathcal{D}_\sigma=\mathcal{D}=2$, which  proves to be the most challenging.

The numerical analysis indicates that the profile of the critical chemical
potential is almost quadratic, but subject to logarithmic corrections.
The ODE to be solved  for $\bar{\mu}\doteq-\hat{\mu}/\hat{T}_\mathrm{H}=-\ln\hat{z}$ reads
\begin{subequations}\label{eq:422}
\begin{align}\label{eq:422a}
& \bar{\mu}''+\frac{1}{\hat{r}}\,\bar{\mu}'
+4\ln\big(1-e^{-\bar{\mu}}\big)=0, \\ \label{eq:422b}
& \bar{\mu}(0)=\bar{\mu}'(0)=0. 
\end{align}
\end{subequations}
The normalization condition for given $\hat{R}$, 
\begin{align}\label{eq:423}
2\hat{T}_\mathrm{H}\int_0^{\hat{R}}d\hat{r}\,\hat{r}
\ln\big(1-e^{-\bar{\mu}}\big)=-1,
\end{align}
determines the critical temperature $\hat{T}_\mathrm{H}$ or vice versa.
The absence of parameters in (\ref{eq:422}) makes the critical singularities independent of $\hat{R}$.

\begin{figure}[b]
  \begin{center}
 \includegraphics[width=55mm]{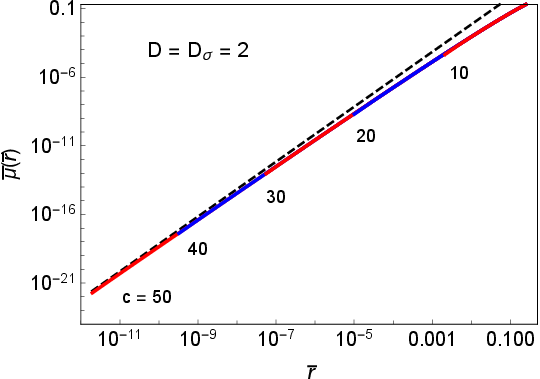}
\end{center}
\caption{Numerical integration of the ODE (\ref{eq:422}) with five different initial values $\hat{r}_0$. 
The initial values are the locations where the color (or shade) of the curve changes.
The initial values are $\hat{r}_0=2.1\times10^{-3}$, $1.0\times10^{-5}$, $5.6\times10^{-8}$, $3.3\times10^{-10}$, and $2.0\times10^{-12}$.
The dashed line represents the function $c\hat{r}^2$ with $c=50$.}
  \label{fig:10}
\end{figure}

The numerical integration cannot be started at ${\hat{r}=0}$.
All terms of (\ref{eq:422a}) diverge.
We circumnavigate this problem by using the ansatz, 
\begin{equation}\label{eq:424}
\bar{\mu}_0(\hat{r})=c\hat{r}^2,
\end{equation}
combined with the insistence that it satisfy (\ref{eq:422a}) at the initial radius $\hat{r}_0>0$ of our choice.
The amplitude $c$ which does the trick depends on $\hat{r}_0$ as follows:
\begin{equation}\label{eq:425}
\hat{r}_0^{-2}=c\,e^c.
\end{equation}
This relation encodes the logarithmic correction to the quadratic profile in a roundabout way. 
As $\hat{r}_0$ is made smaller, the amplitude $c$ increases without bound. 
The real solution has infinite curvature at $\hat{r}=0$.
In Fig.~\ref{fig:10} we show solutions of (\ref{eq:422}) with five initial values $\hat{r}_0$ corresponding to $c=10,20, \ldots, 50$.
All solutions are found to neatly connect in a progression of precision.
The dashed line represents (\ref{eq:424}) with $c=50$.
The critical density inferred from (\ref{eq:19}) is $\rho(\hat{r})=-\hat{T}_\mathrm{H}\ln(1-e^{-\bar{\mu}})$.

\subsection{Salient features}\label{sec:sal-fea-cylin}
Changing the symmetry of the cluster from planar to cylindrical has drastic consequences for both the MB gas and the BE gas.
At high temperature, the BE and MB density profiles look very similar. 
Differences first show up near the center of the cluster, where the density is highest.
The central density of the BE gas initially grows faster than its MB counterpart. 
At $\hat{T}_\mathrm{H}$ the BE central density diverges in $\mathcal{D}=2$ and acquires a cusp singularity in $\mathcal{D}=3$.
In both cases, the singularity is associated with the onset of condensation.
The MB central density, by contrast, diverges at $\hat{T}_\mathrm{MB}$, which is lower than $\hat{T}_\mathrm{H}$ in both $\mathcal{D}=2$ and $\mathcal{D}=3$.

The free energy plotted versus temperature shows features typical of a first-order transition. 
However, the nontrivial mechanical stability condition prevents condensation to take place gradually at a fixed temperature.
Caloric curves give us landmark values for temperature (in the canonical ensemble) or internal energy (in the microcanonical ensemble), where either the mixed or the purely gaseous macrostate becomes unstable. 
The instabilities in the forward and reverse directions happen at different
points on the caloric curve, which is indicative of hysteretic effects.

%
\section{Spherical symmetry}\label{sec:sphere}
%
Density profiles of BE clusters with $\mathcal{D}_\sigma=3$ are analyzed here in
$\mathcal{D}=3$ via 
solutions of Eqs.~(\ref{eq:18})--(\ref{eq:19}) for closed systems of finite mass.
Stable systems without confinement do exist, but they have infinite mass
and are not considered here.
Somewhat different scenarios unfold under confinement with small, intermediate, or large radius $\hat{R}$.

In all three regimes, the BE gas exhibits MB behavior at sufficiently high temperature as expected.
The MB gas in $\mathcal{D}_\sigma=\mathcal{D}=3$ under confinement at radius
$\hat{R}$ is known 
to be stable above the temperature,
\begin{equation}\label{eq:428} 
\hat{T}_\mathrm{C}=\frac{\bar{T}_\mathrm{C}}{\hat{R}},\quad 
\bar{T}_\mathrm{C}=0.794422\ldots, 
\end{equation}
where it suffers a collapse \cite{emden,lbw,aaiso}.
The $\hat{R}$-independent threshold temperature $\bar{T}_\mathrm{C}$ is a consequence of the scale invariance discussed in Sec~\ref{sec:MB-limit}.

Bosonic statistics, known to render stability against gravity more precarious, is expected to initiate condensation at a temperature higher than $\hat{T}_\mathrm{C}$.
However, condensation is not collapse, even though both events are precipitous in this case. 
Recall that condensation was also abrupt in cylindrical BE clusters, yet different from the collapse of MB clusters (Sec.~\ref{sec:cylind}).
In planar BE clusters, by contrast, condensation was found to be gradual and
planar MB clusters do not collapse at all (Sec.~\ref{sec:planar}).

One salient feature of $\mathcal{D}=3$ noted earlier for planar and cylindrical symmetry and relevant here for spherical symmetry is that the density at the center of the cluster remains finite at criticality. 
The critical density profile is expressible as a power series with a negative first derivative, indicative of a cusp singularity, as worked out in Appendix~\ref{sec:appa}.
In the following, we analyze the regime of tight confinement in some detail and then highlight differences realized in the regimes of intermediate and wide confinement.
The three regimes are represented by systems with confining radii $\hat{R}=1,10,100$.

\subsection{Tight confinement}\label{sec:tight-conf}
\begin{figure}[htb]
  \begin{center}
\includegraphics[width=42mm]{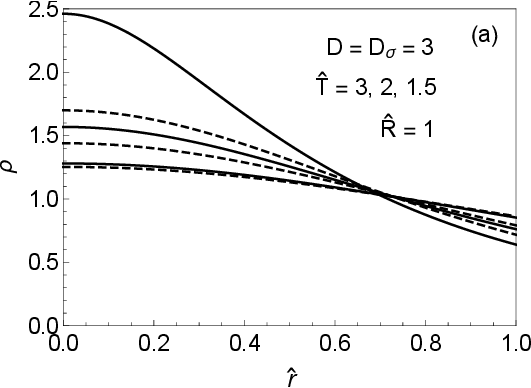}%
\includegraphics[width=42mm]{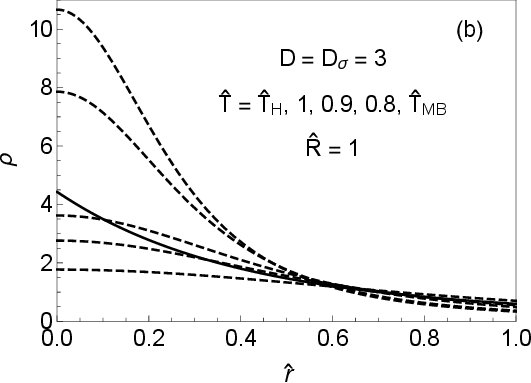}
 \end{center}
\caption{Comparison of BE density profiles (solid lines) and MB density profiles (dashed lines) for $\hat{R}=1$. (a) Demonstration of convergence at high $\hat{T}$ and trend of deviations. (b) Critical BE profile at $\hat{T}_\mathrm{H}=1.42249$ compared with MB profiles at $\hat{T}_\mathrm{H}\leq\hat{T}\leq\hat{T}_\mathrm{C}=0.794422$.}
  \label{fig:11}
\end{figure}

We begin at high temperatures, as we did with clusters of planar and cylindrical symmetry.
In Fig.~\ref{fig:11} we show comparative plots of BE and MB density profiles.
Panel (a) demonstrates the gradual upward deviation of the BE density from the MB profile as $\hat{T}$ is lowered from a high value.
The BE gas is weaker in withstanding gravitational pressure. 
Its density near the center of the cluster rises faster and hits a singularity earlier.
Multiple occupancy of one-particle states is accommodated by BE statistics, ignored by MB statistics, and, as explored in \cite{sgcfd}, prohibited by FD statistics.

The criticality of the BE gas is signalled by the central fugacity reaching the value, $\hat{z}(0)=1$, and by the density profile acquiring a linear cusp singularity.
It happens at 
\begin{equation}\label{eq:429} 
\hat{T}_\mathrm{H}=1.42249\ldots
\end{equation}
The critical BE profile is shown as solid line in panel (b).
The MB profile at $\hat{T}_\mathrm{H}$ is the dashed line with the lowest value at $\hat{r}=0$.
As $\hat{T}$ is lowered from there, the MB central density keeps rising, but only to a finite value.
At the temperature $\hat{T}_\mathrm{C}=0.794422\ldots$,
where the MB cluster collapses gravitationally, its density profile still has
zero initial slope, in strong contrast to MB clusters with cylindrical symmetry,
where the central density diverges at the verge of collapse
(Sec.~\ref{sec:cylind}).

From this first examination we have learned three salient facts: (i) spherical BE clusters behave quite differently from their MB counterparts;
(ii) spherical MB clusters collapse quite differently from their cylindrical counterparts; (iii) spherical BE clusters behave similarly to their cylindrical counterparts (in $\mathcal{D}=3$).  

\begin{figure}[b]
  \begin{center}
\includegraphics[width=42mm]{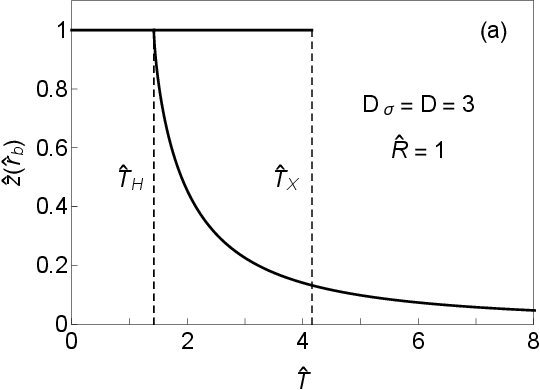}%
\includegraphics[width=42mm]{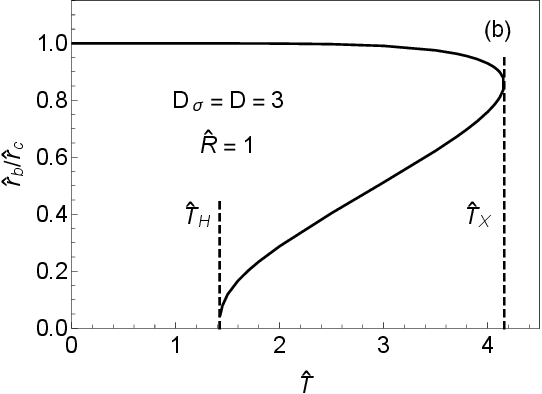}
 \end{center}
\caption{(a) Central or interfacial fugacity. 
Purely gaseous states are represented by points along the curve. 
Points on the horizontal line for $0\leq\hat{T}\leq\hat{T}_\mathrm{X}$ represent mixed-phase states.
(b) Interfacial radius of mixed-phase solutions, which are unique for $\hat{T}<\hat{T}_\mathrm{H}$.
Two solutions exist for $\hat{T}_\mathrm{H}<\hat{T}<\hat{T}_\mathrm{X}$.}
\label{fig:12}
\end{figure}

Figure~\ref{fig:12}(a) shows the central or interfacial fugacity $\hat{z}(\hat{r}_\mathrm{b})$ versus temperature. 
At $\hat{T}>\hat{T}_\mathrm{X}\simeq4.16$, there exists a unique, noncritical gaseous profile and, at $0<\hat{T}<\hat{T}_\mathrm{H}$, a unique mixed-phase profile composed of a BEC core and a gaseous halo.
In between, the ODE has three solutions: one is a pure gas state with $\hat{z}(0)<1$ and the other two are mixed-phase states, which have $\hat{z}(\hat{r}_\mathrm{b})=1$.
The relevant boundary conditions are the (critical) interface fugacity,
\begin{subequations}\label{eq:435}
\begin{align}\label{eq:435a}
\hat{z}(\hat{r}_\mathrm{b})=1,
\end{align}
the slope required for mechanical stability,
\begin{align}\label{eq:435b}
\hat{z}'(\hat{r}_\mathrm{b})=
 -\frac{2}{\hat{T}}\frac{\hat{r}_\mathrm{b}}{\hat{r}_\mathrm{c}^3}\,
 \hat{z}(\hat{r}_\mathrm{b}), \quad
  \left(\frac{\hat{r}_\mathrm{b}}{\hat{r}_\mathrm{c}}\right)^3
 =1-\frac{N_\mathrm{gas}}{N},
\end{align}
and the BEC radius $\hat{r}_\mathrm{b}$ to enforce mass conservation,
\begin{align}\label{eq:435c}
 3\hat{T}^{3/2}\int_{\hat{r}_\mathrm{b}}^{\hat{R}} d\hat{r}
\,\hat{r}^2g_{3/2}(\hat{z})=\frac{N_\mathrm{gas}}{N}.
\end{align}
\end{subequations}

The choice of reference radius $\hat{r}_\mathrm{c}=0.1$ is large enough to accommodate short-distance regularization computationally, yet sufficiently small to permit a realistic physical representation of the gaseous halo.
Figure~\ref{fig:12}(b) shows the variation of $\hat{r}_\mathrm{b}$ across the range $0\leq\hat{T}\leq\hat{T}_\mathrm{X}$, where mixed-phase solutions exist.
One or two solutions exist depending on whether $\hat{T}$ falls below or above $\hat{T}_\mathrm{H}$.
The data used for Fig.~\ref{fig:12}(b) in conjunction with Eq.~(\ref{eq:435b}) produce data for $N_\mathrm{gas}/N$ versus $\hat{T}$, not shown but of a shape similar to Fig.~\ref{fig:5}(b) apart from the low-$T$ asymptotics, which now is $\sim\hat{T}^{5/2}$.

\begin{figure}[htb]
\begin{center}
\includegraphics[width=42.5mm]{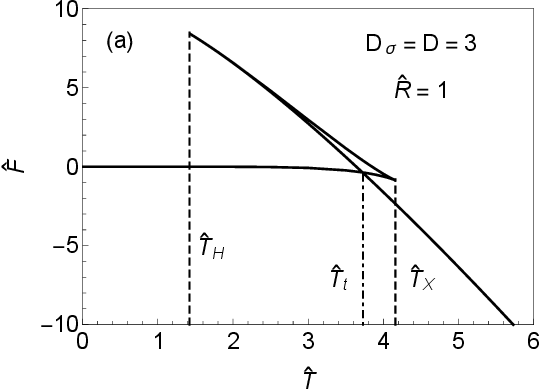}%
\includegraphics[width=41.5mm]{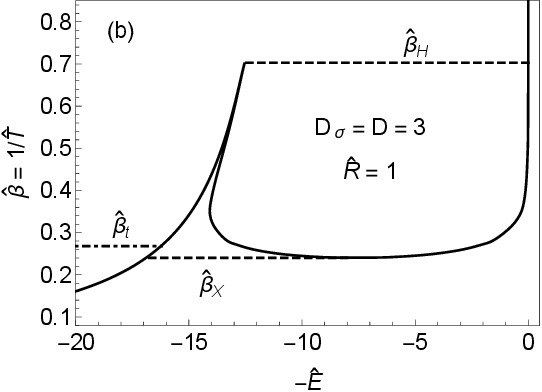}
\end{center}
\caption{(a) Helmholtz free energy, $\hat{{F}}$, plotted versus
scaled temperature and (b) caloric curve for a cluster confined to radius
$\hat{R} = 1$.}
\label{fig:13}
\end{figure}

The conditions of mechanical stability and thermal equilibrium can be read off
the free energy plot and caloric curve shown in Fig.~\ref{fig:13}.
Note the similarity to Fig.~\ref{fig:6} for cylindrical clusters.
The macrostate with lowest free energy is a pure gas at $\hat{T}>\hat{T}_\mathrm{t}$ and of mixed-phase at $\hat{T}<\hat{T}_\mathrm{t}$.
The pure-gas state at $\hat{T}_\mathrm{H}<\hat{T}<\hat{T}_\mathrm{t}$ and the mixed-phase state at $\hat{T}_\mathrm{t}<\hat{T}<\hat{T}_\mathrm{X}$ are metastable.
The highest branch of free energy, curved upward, is unstable.

The shape of the caloric curve shows that the self-crossing at $\hat{T}_\mathrm{t}$ does not have the significance for the phase behavior it would have in a homogeneous system. 
According to the Poincar\'e conditions for the canonical ensemble, the spherical cluster stays, upon cooling, in the gas phase down past $\hat{T}_\mathrm{t}$ to $\hat{T}_\mathrm{H}$.
At this point, a mechanical instability occurs, which brings it to a thermal equilibrium in a mixed-phase state at the same temperature, displaced horizontally to the right in Fig.~\ref{fig:13}(b).
Conversely, when the system in the mixed-phase state is heated up, nothing
dramatic happens at $\hat{T}_\mathrm{t}$, but it suffers a mechanical
instability at $\hat{T}_\mathrm{X}$ and settles in a gas state at that
temperature, displaced horizontally to the left in Fig.~\ref{fig:13}(b).

There is no obvious isothermal sequence of mechanically stable macrostates that connects the gaseous equilibrium macrostates at $\hat{T}>\hat{T}_\mathrm{t}$ and the mixed-phase macrostates at $\hat{T}<\hat{T}_\mathrm{t}$. 
The nontrivial nature of mechanical stability in self-gravitating clusters  makes it unlikely to exist.
Nevertheless, the transition as encoded in the curves of Fig.~\ref{fig:13} for the canonical ensemble is discontinuous in the sense that the order parameter vanishes abruptly at $\hat{T}_\mathrm{X}$ on the way up in temperature and jumps to a nonzero value at $\hat{T}_\mathrm{H}$ on the way down.
Effects of hysteresis are an intrinsic feature of this transition.

\begin{figure}[htb]
\begin{center}
\includegraphics[width=42mm]{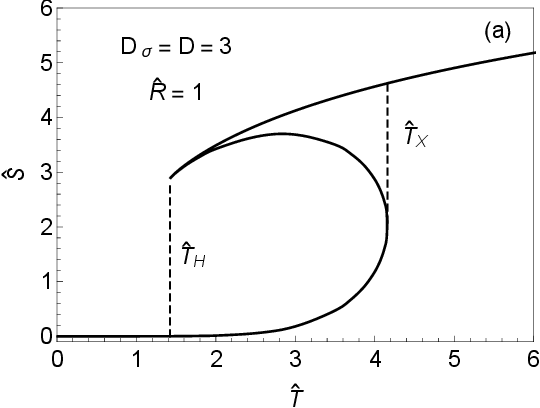}%
\includegraphics[width=42mm]{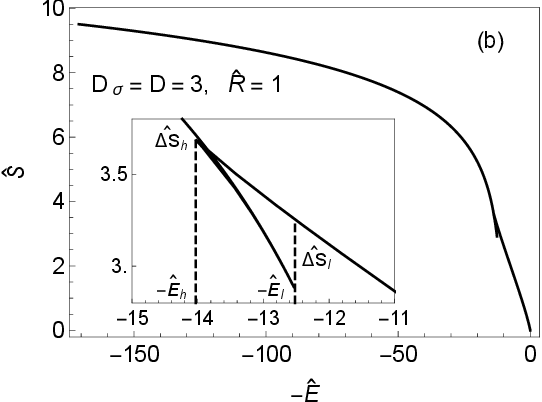}
\end{center}
\caption{Scaled entropy plotted (a) versus scaled temperature and (b) versus scaled internal energy for a cluster confined to radius $\hat{R} = 1$.}
\label{fig:14}
\end{figure}

The instabilities at $\hat{T}_\mathrm{H}$ and $\hat{T}_\mathrm{X}$ leave their characteristic signatures also in the entropy plot of Fig.~\ref{fig:14}(a).
Here they begin at points of infinite slope and proceed along vertical lines. 
On the way to lower temperature, the instability (at $\hat{T}_\mathrm{H}$) requires extraction of entropy (or heat) in order to maintain the same $\hat{T}$.
On the way up in temperature, the instability (at $\hat{T}_\mathrm{X}$) requires heat to be added for the same purpose.
While heat is extracted at $\hat{T}_\mathrm{H}$ and added at
$\hat{T}_\mathrm{X}$, both instabilities are associated with a decrease in free
energy [Fig.~\ref{fig:13}(a)].

In an astrophysical context, the relatively fast processes triggered by the mechanical instability are more appropriately described within the microcanonical ensemble.
There is little opportunity for heat exchange during the time it takes the cluster to settle down in a new macrostate in the wake of a mechanical instability.
The thermodynamic equilibrium state is then identified as the one with the highest entropy when plotted versus energy as in Fig.~\ref{fig:14}(b).
Mechanical instabilities now occur at $\hat{E}_\mathrm{l}$ on the way down and at $\hat{E}_\mathrm{h}$ on the way up in energy.
In both instances the instability is associated with an entropy increase, $\Delta\hat{S}_\mathrm{l}$ and $\Delta\hat{S}_\mathrm{h}$, respectively.

\begin{figure}[htb]
  \begin{center}
 \includegraphics[width=41mm]{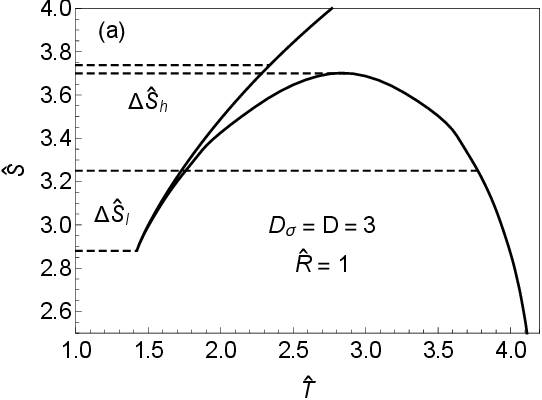}%
 \includegraphics[width=43mm]{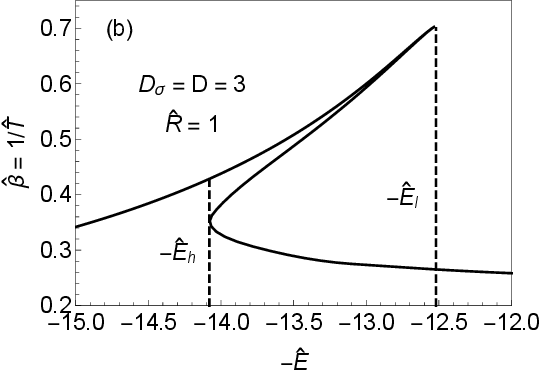}
 \end{center}
\caption{Partial views of (a) entropy versus temperature and (b) caloric curve.
Data previously used in Fig.~\ref{fig:13}(b) and \ref{fig:14}(a) are reproduced
within zoomed-in window. }
  \label{fig:15}
\end{figure}

It is instructive to compare the endpoints of the mechanical instabilities in
the two ensembles as is done in the entropy plots of Figs.~\ref{fig:14}(a) and 
\ref{fig:15}(a) as well as 
in the caloric curves depicted in Figs.~\ref{fig:13}(b) and \ref{fig:15}(b).
In the microcanonical ensemble, the instability occurs at a local maximum or
minimum of $\hat{S}$ when plotted versus $\hat{T}$  and at points of infinite
slope of the caloric curve as identified in panels (a) and (b) of
Fig.~\ref{fig:15}, respectively.
The end state at $\hat{E}_\mathrm{l}$ ($\hat{E}_\mathrm{h}$) has higher (lower) temperature, but both end states have higher entropy.

In the canonical ensemble, by contrast, the instability occurs at points where
$\hat{S}$ has infinite slope when plotted versus $\hat{T}$
[Fig.~\ref{fig:14}(a)] and at points of zero slope in the caloric curve
[Fig.~\ref{fig:13}(b)].
The end state at $\hat{T}_\mathrm{X}$ ($\hat{T}_\mathrm{H}$) has higher (lower) entropy, but both end states have lower free energy.
The points of instability are, quite generally, closer together in the microcanonical ensemble, as already seen for cylindrical clusters, but the spike in the caloric curves puts one instability to almost identical locations in the two ensembles.

\subsection{Intermediate confinement}\label{sec:inter-conf}
\begin{figure}[htb]
  \begin{center}
\includegraphics[width=42mm]{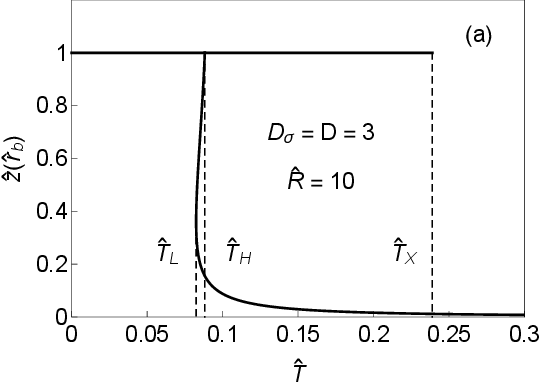}%
\includegraphics[width=42mm]{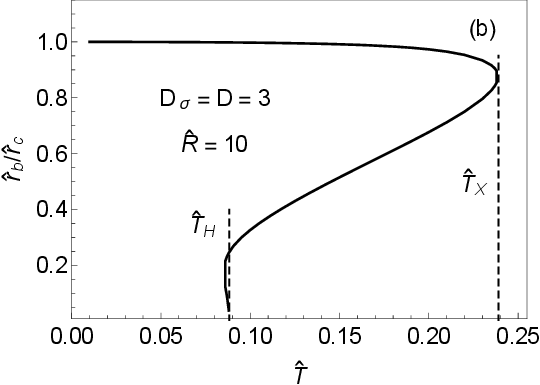}
\includegraphics[width=42mm]{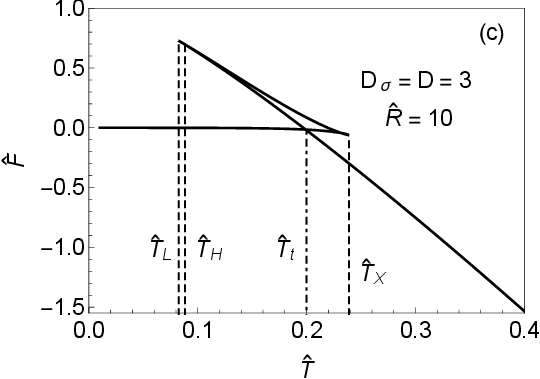}%
\includegraphics[width=42mm]{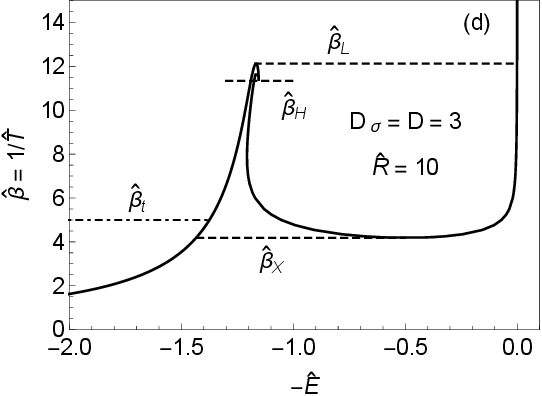}
 \end{center}
\caption{(a) Central or interfacial fugacity, (b) interfacial radius of mixed-phase solutions, (c) free energy, and (d) caloric curve.}
\label{fig:16}
\end{figure}

\begin{figure}[t]
\begin{center}
\includegraphics[width=42mm]{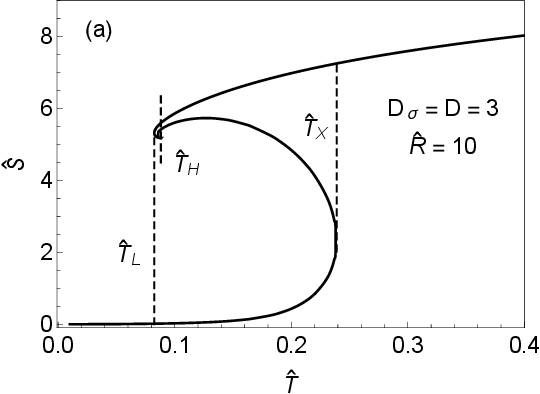}%
\includegraphics[width=42mm]{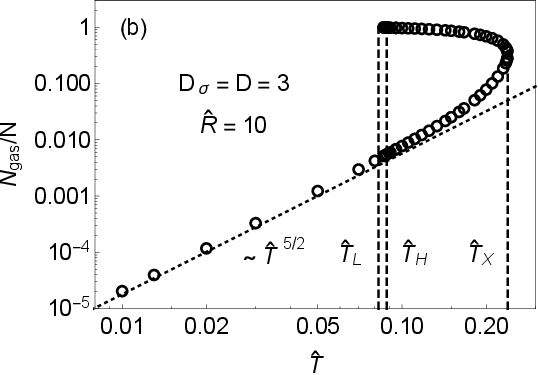}
\includegraphics[width=42mm]{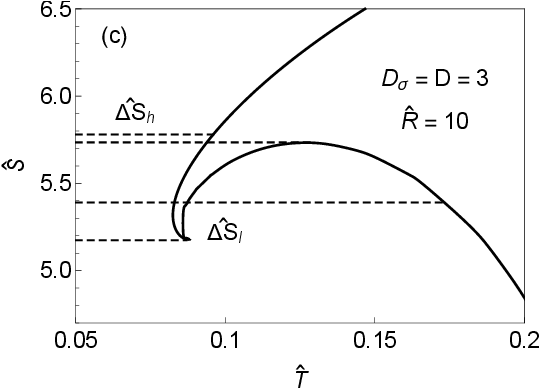}%
\includegraphics[width=42mm]{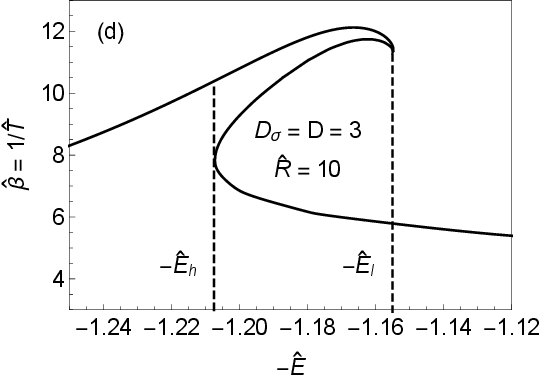}
\end{center}
\caption{(a) Entropy, (b) fraction of particles in the gas phase, (c) zoomed-in
detail of entropy curve, (d) zoomed-in detail of caloric curve. }
\label{fig:17}
\end{figure}

When increasing the confining radius of the cluster to an intermediate radius of confinement, $1.64\lesssim \hat{R} \lesssim 27.35$, new features appear. 
All results presented here are for $\hat{R}=10$ and we use $\hat{r}_\mathrm{c}=1$ for the reference radius throughout.
Each panel of Figs.~\ref{fig:16} and \ref{fig:17} with one exception have counterparts in Sec.~\ref{sec:tight-conf} for direct comparison.

One principal change caused by the widening of confinement is that the macrostate of pure gas at the lowest temperature is no longer critical [Fig.~\ref{fig:16}(a)].
This brings an additional landmark temperature into play: $\hat{T}_\mathrm{L}\simeq0.0825$.
It is only slightly lower than the temperature $\hat{T}_\mathrm{H}\simeq0.0882$, where the gaseous macrostate reaches criticality at the center of the cluster and acquires a linear cusp singularity in its density profile.
The highest temperature $\hat{T}_\mathrm{X}\simeq0.239$, for which a mixed-phase state with a BEC core and a gaseous halo coexist is much higher.

The horizontal line in Fig.~\ref{fig:16}(a) represents a unique mixed-phase
state at $\hat{T}<\hat{T}_\mathrm{H}$ and a pair of such states at
$\hat{T}>\hat{T}_\mathrm{H}$, one 
stable and the other unstable (in the canonical ensemble).
Panel (b) shows the position of the interface as a function of $\hat{T}$ of these mixed-phase states. 
The functional dependence is similar to what we have seen under tight confinement, except for the hint of an additional solution near $\hat{T}_\mathrm{H}$.
The free energy plot of panel (c) looks similar as well, except for an
additional unstable branch (not resolved) near the
uppermost tip.

The temperature $\hat{T}_\mathrm{t}$ at the border between the mixed-phase and a purely gaseous equilibrium macrostates, has again no significance in the actual phase transitions due to the nontrivial mechanical stability condition.
The evidence is shown in the caloric curve of panel (d).
When the cluster is quasi-statically cooled down from high temperature, it becomes mechanically unstable at $\hat{T}_\mathrm{L}$, sheds heat to the environment while condensing, and settles in the mixed-phase state where the highest dashed line intersects the curve on the right.

Conversely, when the mixed-phase cluster is heated up it undergoes a mechanical instability at the much higher temperature $\hat{T}_\mathrm{X}$, absorbs heat while the condensate evaporates, and settles in the gaseous state where the lowest dashed line intersects the curve on the left.
The additional feature of the caloric curve near $\hat{T}_\mathrm{H}$ in comparison to its tight-confinement counterpart [Fig.~\ref{fig:13}(b)] has no bearing on the phase behavior just described.

The heat absorption at $\hat{T}_\mathrm{X}$ and heat expulsion at $\hat{T}_\mathrm{L}$ are also evident in the entropy plot of Fig.~\ref{fig:17}(a).
Again the additional feature of the entropy curve near $\hat{T}_\mathrm{H}$, not present in its tight-confinement counterpart [Fig.~\ref{fig:14}(a)], makes no difference.
In Fig.~\ref{fig:17}(b) we show data for one aspect of the phase behavior representative of all three regimes, namely the low-$\hat{T}$ asymptotics of the order parameter. 
The power-law asymptotics, $N_\mathrm{gas}/N\sim\hat{T}^{\mathcal{D}/2+1}$, mentioned in the context of (\ref{eq:352}) is borne out convincingly by the data, as it is in the other two regimes (not shown). 
The data of the overhanging branch pertain to mechanically unstable two-phase macrostates.

Finally, panels (c) and (d) illustrate the transitions as they unfold in the microcanonical ensemble, where no heat transfer with the environment takes place.
Here the mechanical instabilities proceed at different values of constant energy. 
Unlike in the canonical ensemble, both instabilites are associated with an increase in entropy, which here plays the role of thermodynamic potential.

\subsection{Wide confinement}\label{sec:wide-conf}
\begin{figure}[t]
  \begin{center}
\includegraphics[width=42mm]{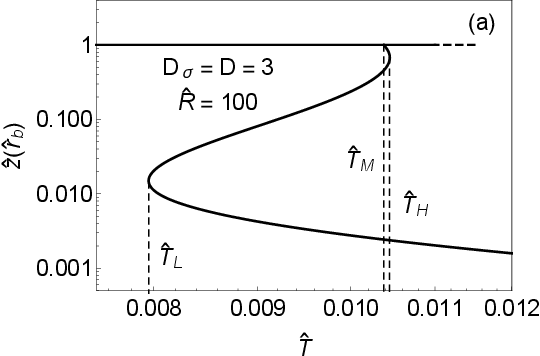}%
\includegraphics[width=42mm]{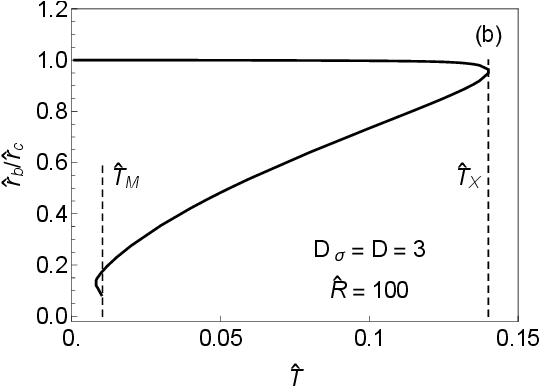}
\includegraphics[width=42mm]{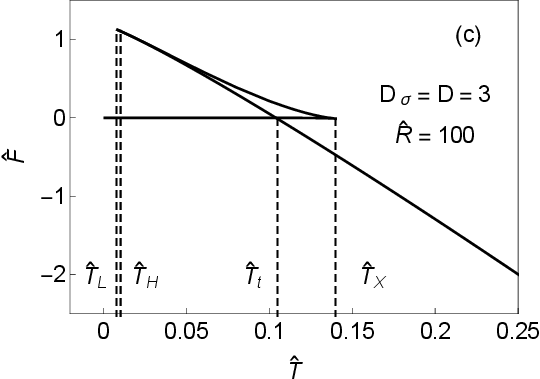}%
\includegraphics[width=42mm]{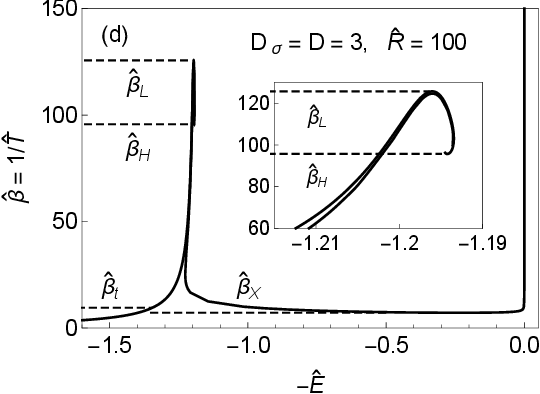}
 \end{center}
\caption{(a) Central or interfacial fugacity, (b) interfacial radius of mixed-phase solutions, (c) free energy, and (d) caloric curve.}
\label{fig:18}
\end{figure}

When the radius of confinement is relaxed beyond the intermediate regime, a new landmark temperature enters the game, but with very limited impact.
The new features are summarized in Fig.~\ref{fig:18} for the representative case $\hat{R}=100$ and $\hat{r}_\mathrm{c}=1$.
There is now a narrow temperature interval, where three gaseous macrostates 
coexist. 
The highest temperature, $\hat{T}_\mathrm{X}\simeq0.14$ is now far higher than the temperatures $\hat{T}_\mathrm{L}\simeq0.00795$ and $\hat{T}_\mathrm{H}\simeq0.01045$ which delimit the range of multiple gaseous macrostates or the temperature $\hat{T}_\mathrm{M}\simeq0.01038$, where the gaseous macrostate becomes critical.
Additional (unstable) gaseous macrostates come into play as $\hat{R}$ is increased further.

\begin{figure}[b]
\begin{center}
\includegraphics[width=42mm]{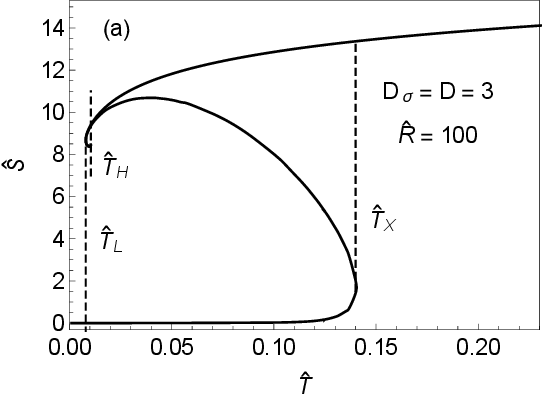}%
\includegraphics[width=42mm]{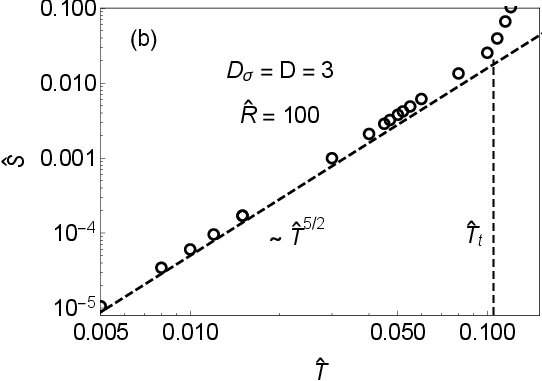}
\includegraphics[width=42mm]{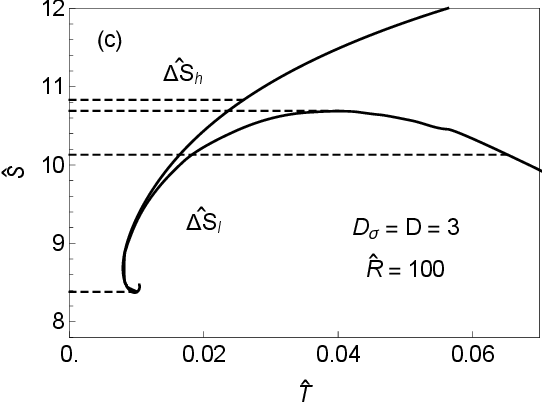}%
\includegraphics[width=42mm]{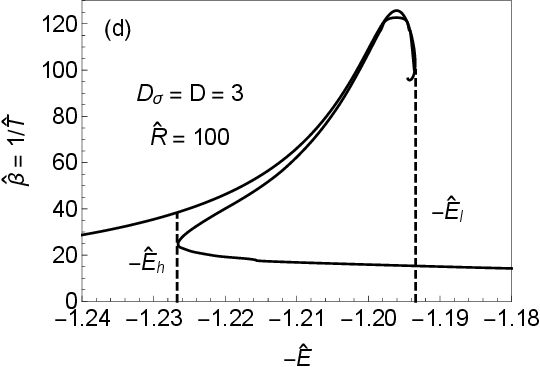}
\end{center}
\caption{(a) Entropy, (b) low-$\hat{T}$ asymptotics of entropy, (c) zoomed-in
detail of entropy, (d) zoomed-in detail of caloric curve. }
\label{fig:19}
\end{figure}

The free energy curve has an additional fold near the highest tip, unresolved in panel (c). 
The caloric curve further bends into an incipient spiral.
Finally, Fig.~\ref{fig:19}(b) is representative of all three regimes as was  Fig.~\ref{fig:17}(b). 
It shows the low-$\hat{T}$ asymptotics of the entropy. 
The power-law behavior was predicted earlier in (\ref{eq:352}).
None of these features have any qualitative impact on the phase behavior, as illustrated in Fig.~\ref{fig:19}. 
In the canonical ensemble, a mechanical instability at $\hat{T}_\mathrm{L}$ on the way down in temperature triggers an abrupt onset of condensation. On the way up in temperature, the mixed-phase state disappears equally abruptly in a mechanical instability at $\hat{T}_\mathrm{X}$.

In situations where the microcanonical ensemble provides a more realistic description, transitions occur at energies $\hat{E}_\mathrm{l}$ and $\hat{E}_\mathrm{h}$.
In both ensembles, the instabilities trigger processes that move the relevant thermodynamic potential toward equilibrium -- the Helmoltz free energy $\hat{F}$ toward a newly available minimum in the canonical ensemble and the entropy $\hat{S}$ toward a newly available maximum in the microcanonical ensemble.

%
\section{Conclusion and outlook}\label{sec:concl}
%
Self-gravitating clusters of bosonic particles initiate condensation in processes that depend on the symmetry of the cluster and the dimensionality of the space. 
Mechanical instabilities play a key role in some cases and produce effects of hysteresis.
The focus of this work has been on the density profiles of gaseous clusters  and of gaseous halos in mixed-phase clusters.
For that purpose, we have worked with a provisional BEC in the form of a highly compact reference state.

The results of this work, specifically the interfacial pressure, are a prerequisite for the analysis of BEC-core density profiles, whose shapes must be investigated on a different length scale. 
The results of that analysis can then be used to modify the provisional BEC into a more realistic shape for an overall improvement of mixed-state profiles.
The free energy expressions and the interface boundary conditions in
Sec.~\ref{sec:fund} are designed for adaption to this purpose.

In a companion paper \cite{sgcfd}, we have analyzed self-gravitating FD clusters of the same symmetries and in spaces of the same dimensionalities. 
As expected, FD and BE clusters evolve differently upon cooling from their common low-density MB limit.
Whereas FD clusters tend to be amenable to exact analysis in the fully
degenerate limit, BE clusters tend to facilitate exact results at criticality.

Emerging from these parallel studies is a curious correspondence in the phase behavior of spherical BE and FD clusters. 
Both quantum statistics give rise to abrupt changes associated with mechanical instabilities and hysteretic effects.
In the BE clusters described here, the instabilities trigger processes between purely gaseous profiles and mixed-state profiles of a BEC core and a gaseous halo.
In the FD clusters described in \cite{sgcfd}, by contrast, the instabilities initiate processes between gaseous states with different density profiles and different degrees of degeneracy.

Unanswered in both studies is the question about the existence and nature of a quasi-static process that links the equilibrium states on either side of the pair of mechanical instabilities. 
In the FD study we have suggested that the answer is to be looked for in macrostates of phase coexistence between gaseous states of different profiles. 
In the case of BE cluster, the path toward an answer is more complicated and will have to await the calculation of BEC density profiles under the weight of surrounding halos such as have emerged from this study.

\appendix

%
\section{Length scale and energy scale}\label{sec:appb}
%
The scales $r_\mathrm{s}$ and $k_\mathrm{B}T_\mathrm{s}$ inferred from (\ref{eq:16}) and (\ref{eq:17}) are 
\begin{equation}\label{eq:203}
r_\mathrm{s}^{\mathcal{D}_\sigma}=
\frac{\mathcal{D}_\sigma}{\mathcal{A}_{\mathcal{D}_\sigma}}
(2\pi\hbar^2)^{\mathcal{D}/2}\tilde{N}\,m^{-\mathcal{D}/2}
(k_\mathrm{B}T_\mathrm{s})^{-\mathcal{D}/2},
\end{equation}
\begin{align}\label{eq:202}
& (k_\mathrm{B}T_\mathrm{s})^{1+\mathcal{D}/\mathcal{D}_\sigma-\mathcal{D}/2}
=\frac{1}{2}\frac{\mathcal{A_D}}{\mathcal{D}_\sigma}
\left(\frac{\mathcal{A}_{\mathcal{D}_\sigma}}{\mathcal{D}_\sigma}\right)^{-2/\mathcal{D}_\sigma}  \\ \rule[0mm]{-2mm}{6mm}
& \hspace{5mm}\times\,G_\mathcal{D}\,
(2\pi\hbar^2)^{\mathcal{D}/\mathcal{D}_\sigma-\mathcal{D}/2} 
m^{2-\mathcal{D}/\mathcal{D}_\sigma+\mathcal{D}/2}
\tilde{N}^{2/\mathcal{D}_\sigma}.\nonumber 
\end{align}
The results for the cases considered in this work, expressed as functions of $N$ (number of particles) and $m$ (particle mass) and as function of $m$ and $\tilde{m}_\mathrm{tot}$ simplify as listed below. 
The unit of the gravitational constant $G_\mathcal{D}$ depends on $\mathcal{D}$. Only $G_3$, of course, is known and relevant for astrophysics.
\begin{itemize}

\item[$\rhd$] $\mathcal{D}_\sigma=1,~ \mathcal{D}=1$:
\begin{equation*}\label{eq:204}
k_\mathrm{B}T_\mathrm{s}=\frac{1}{2}\pi^{1/3}(\hbar G_1)^{2/3}mN^{4/3}
=\frac{1}{2}\pi^{1/3}(\hbar G_1)^{2/3}\frac{m_\mathrm{tot}^{4/3}}{m^{1/3}},
\end{equation*}
\begin{equation}\label{eq:205}
r_\mathrm{s}=\left(\frac{\pi\hbar^2}{G_1}\right)^{1/3}\frac{N^{1/3}}{m}
=\left(\frac{\pi\hbar^2}{G_1}\right)^{1/3}\frac{m_\mathrm{tot}^{1/3}}{m^{4/3}};
\end{equation}

\item[$\rhd$] $\mathcal{D}_\sigma=1,~ \mathcal{D}=2$:
\begin{equation*}\label{eq:206}
k_\mathrm{B}T_\mathrm{s}=\pi\hbar\sqrt{\frac{G_2}{2}}m^{1/2}\tilde{N}
=\pi\hbar\sqrt{\frac{G_2}{2}}\frac{\tilde{m}_\mathrm{tot}}{m^{1/2}},
\end{equation*}
\begin{equation}\label{eq:207}
r_\mathrm{s}=\hbar\sqrt{\frac{2}{G_2}}m^{-3/2};
\end{equation}

\item[$\rhd$] $\mathcal{D}_\sigma=1,~ \mathcal{D}=3$:
\begin{equation*}\label{eq:208}
k_\mathrm{B}T_\mathrm{s}=\pi(2\hbar^6G_3^2)^{1/5}m^{1/5}\tilde{N}^{4/5}
=\pi(2\hbar^6G_3^2)^{1/5}\frac{\tilde{m}_\mathrm{tot}^{4/5}}{m^{3/5}},
\end{equation*}
\begin{equation}\label{eq:209}
r_\mathrm{s}=\left(\frac{2\hbar^6}{G_3^3}\right)^{1/5}
\!\!\!m^{-9/5}\tilde{N}^{-1/5}
=\left(\frac{2\hbar^6}{G_3^3}\right)^{1/5}
\!\!\!m^{-8/5}\tilde{m}_\mathrm{tot}^{-1/5};
\end{equation}

\item[$\rhd$] $\mathcal{D}_\sigma=2,~ \mathcal{D}=2$:
\begin{equation}\label{eq:210}
k_\mathrm{B}T_\mathrm{s}=\frac{G_2}{2}m^2N=\frac{G_2}{2}m\,m_\mathrm{tot},
\quad r_\mathrm{s}=\frac{2\hbar}{\sqrt{G_2}}m^{-3/2};
\end{equation}

\item[$\rhd$] $\mathcal{D}_\sigma=2,~ \mathcal{D}=3$:
\begin{equation}\label{eq:212}
k_\mathrm{B}T_\mathrm{s}=G_3m^2\tilde{N}=G_3m\,\tilde{m}_\mathrm{tot},
\end{equation}
\begin{equation*}\label{eq:213}
r_\mathrm{s}=\left(\frac{8\pi\hbar^6}{G_3^3}\right)^{1/4}
\!\!\!m^{-9/4}\tilde{N}^{-1/4}
=\left(\frac{8\pi\hbar^6}{G_3^3}\right)^{1/4}
\!\!\!m^{-2}\tilde{m_\mathrm{tot}}^{-1/4};
\end{equation*}

\item[$\rhd$] $\mathcal{D}_\sigma=3,~ \mathcal{D}=3$:
\begin{align}\label{eq:214}
k_\mathrm{B}T_\mathrm{s} &=\frac{G_3^2}{2\hbar^2}(36\pi)^{-1/3}m^5N^{4/3}
\nonumber \\
&=\frac{G_3^2}{2\hbar^2}(36\pi)^{-1/3}m^{11/3}m_\mathrm{tot}^{4/3},
\end{align}
\begin{equation*}\label{eq:215}
r_\mathrm{s}=(36\pi)^{1/3}\frac{\hbar^2}{G_3}m^{-3}N^{-1/3}
=(36\pi)^{1/3}\frac{\hbar^2}{G_3}m^{-8/3}m_\mathrm{tot}^{-1/3}.
\end{equation*}

\end{itemize}

%
\section{Series expansion of critical profiles}\label{sec:appa}
%
For BE clusters in $\mathcal{D}=3$, the profile of the chemical potential can be expanded into a power series beginning with the quadratic term. 
Here we carry out the analysis for a clusters with spherical symmetry $(\mathcal{D}_\sigma=3)$. Results for planar symmetry $(\mathcal{D}_\sigma=1)$ are shown in Sec.~\ref{sec:pla-D3} and for cylindrical symmetry $(\mathcal{D}_\sigma=2)$ at the end of this Appendix. 

Writing $\hat{z}=e^{\hat{\mu}/\hat{T}_\mathrm{c}}$ and setting $\bar{\mu}\doteq-\hat{\mu}/\hat{T}_\mathrm{c}=-\ln\hat{z}$, the rescaled critical chemical potential $\bar{\mu}(\hat{r})$ is the  solution of the following set of equations (for $\mathcal{D}_\sigma=3$):
\begin{subequations}\label{eq:402}
\begin{align}\label{eq:402a}
& \bar{\mu}''+\frac{2}{\hat{r}}\,\bar{\mu}'
-6\hat{T}_\mathrm{c}^{1/2}g_{3/2}(e^{-\bar{\mu}})=0, \\ \label{eq:402b}
& \bar{\mu}(0)=\bar{\mu}'(0)=0. 
\end{align}
\end{subequations}
The value of $\hat{T}_\mathrm{c}$ (named $\hat{T}_\mathrm{H}$ in Secs.~\ref{sec:cylind} and \ref{sec:sphere}) for a given $\hat{R}$ is determined by,
\begin{align}\label{eq:403}
3\hat{T}_\mathrm{c}^{3/2}\int_0^{\hat{R}}d\hat{r}\,\hat{r}^2
g_{3/2}(e^{-\bar{\mu}})=1.
\end{align}
The ODE (\ref{eq:402a}) can be satisfied by a chemical potential expressed as the power series,
\begin{equation}\label{eq:404}
\bar{\mu}(\hat{r})=\sum_{n=2}^\infty a_n\hat{r}^n.
\end{equation}
The boundary conditions (\ref{eq:402b}) are satisfied by construction.
The first two terms of (\ref{eq:402a}) yield the series,
\begin{equation}\label{eq:405}
\bar{\mu}''+\frac{2}{\hat{r}}\,\bar{\mu}'=
\sum_{n=2}^\infty n(n+1)a_n\hat{r}^{n-2}=6a_2+12a_3\hat{r}+20\hat{r}^2+\ldots
\end{equation}
For the last term of (\ref{eq:402a}) we use asymptotic expansion of BE functions:
\begin{align}\label{eq:406}
g_{3/2}\big(e^{-\bar{\mu}}\big)
= \zeta\left(\textstyle\frac{3}{2}\right)-2\sqrt{\pi}\,\bar{\mu}^{1/2}
+\sum_{\ell=1}^\infty\frac{(-1)^\ell}{\ell!}
\zeta\left(\textstyle\frac{3}{2}-\ell\right)\bar{\mu}^\ell.
\end{align}
For the second term we substitute (\ref{eq:404}) and expand binomially:
\begin{align}\label{eq:407}
\bar{\mu}^{1/2} &=\hat{r}\,\sqrt{\sum_{n=2}^\infty a_n\hat{r}^{n-2}}
=\sqrt{a_2}\,\hat{r}\,\sqrt{1+\sum_{n=3}^\infty
\frac{a_{n}}{a_2}\hat{r}^{n-2}} \nonumber \\
&=\sqrt{a_2}\,\hat{r}+\sum_{m=2}^\infty b_m\hat{r}^m,
\end{align}
\begin{align}\label{eq:408}
& b_2=\frac{a_3}{2a_2^{1/2}},\quad
b_3=-\frac{a_3^2-4a_2a_4}{8a_2^{3/2}}, \nonumber \\
& b_4=\frac{a_3^3-4a_2a_3a_4+8a_2^2a_5}{16a_2^{5/2}}, ~\ldots 
\end{align}
By substitution of these expansions into (\ref{eq:402a}), we can determine the expansion coefficients sequentially via the solution of sets of linear equations. 
The first few coefficients thus extracted are 
\begin{align}\label{eq:409}
& a_2(\hat{T}_\mathrm{c})=\hat{T}_\mathrm{c}^{1/2}\, \zeta \left(\textstyle \frac{3}{2}\right), 
\quad a_3(\hat{T}_\mathrm{c})=-\hat{T}_\mathrm{c}^{3/4} 
\sqrt{\pi  \zeta \left(\textstyle \frac{3}{2}\right)}, \nonumber \\
& a_4(\hat{T}_\mathrm{c})={\textstyle \frac{3}{10} \hat{T}_\mathrm{c} \Big[\pi -\zeta \left(\textstyle\frac{1}{2}\right) \zeta
   \left(\textstyle\frac{3}{2}\right)\Big]},  \\ \nonumber 
& a_5(\hat{T}_\mathrm{c})= {\textstyle \frac{1}{100} \hat{T}_\mathrm{c}^{1/4} \sqrt{\frac{\pi }{\zeta \left(\frac{3}{2}\right)}} \Big[26\,\hat{T}_\mathrm{c} \,\zeta
   \left(\textstyle\frac{1}{2}\right) \zeta \left(\textstyle\frac{3}{2}\right)-\pi\,\hat{T}_\mathrm{c}+20\Big]}.
\end{align}
The numerical values of all coefficients depend on the value of $\hat{T}_\mathrm{c}$, which is to be determined numerically via the normalization condition (\ref{eq:403}).
The density profile (\ref{eq:19}) can be expanded into a power series of the form:
\begin{align}\label{eq:410}
\rho(\hat{r}) = \sum_{n=0}^\infty c_n\hat{r}^n.
\end{align}
The first three coefficients read
\begin{align}\label{eq:411}
& c_0=\hat{T}_\mathrm{c}^{3/2} \zeta \left(\textstyle\frac{3}{2}\right),
\nonumber \\
& c_1=-2 \sqrt{\pi } a_2^{1/2} \hat{T}_\mathrm{c}^{3/2}
=-2 \hat{T}_\mathrm{c}^{7/4} \sqrt{\pi  
\zeta \left(\textstyle\frac{3}{2}\right)},
\\ \nonumber 
& c_2= -\frac{\hat{T}_\mathrm{c}^{3/2} \left[a_2^{3/2} \zeta
   \left(\frac{1}{2}\right)+\sqrt{\pi } a_3\right]}{a_2^{1/2}}
   =\hat{T}_\mathrm{c}^2 
   \left[\pi -\zeta \left(\textstyle\frac{1}{2}\right) \zeta
   \left(\textstyle\frac{3}{2}\right)\right].
\end{align}
Equivalent results for clusters with cylindrical symmetry $(\mathcal{D}_\sigma=2$) look similar:
\begin{align}\label{eq:415}
& \bar{\mu}(\hat{r})=\sum_{n=2}^\infty a_n\hat{r}^n, \nonumber \\
& a_2(\hat{T}_\mathrm{c})=\hat{T}_\mathrm{c}^{1/2}\, \zeta \left(\textstyle \frac{3}{2}\right), 
\quad a_3(\hat{T}_\mathrm{c})=-\textstyle \frac{8}{9}\,\hat{T}_\mathrm{c}^{3/4} 
\sqrt{\pi  \zeta \left(\textstyle \frac{3}{2}\right)}, \nonumber \\
& a_4(\hat{T}_\mathrm{c})={\textstyle \frac{1}{36} \hat{T}_\mathrm{c} \Big[8\pi -9\zeta \left(\textstyle\frac{1}{2}\right) \zeta
   \left(\textstyle\frac{3}{2}\right)\Big]}, \nonumber \\
& a_5(\hat{T}_\mathrm{c})= {\textstyle \frac{1}{2025}} \hat{T}_\mathrm{c}^{5/4} \sqrt{\frac{\pi }{\zeta \left(\frac{3}{2}\right)}} \nonumber \\
& \hspace{20mm}\times \Big[369\,\hat{T}_\mathrm{c} \,\zeta
   \left(\textstyle\frac{1}{2}\right) \zeta \left(\textstyle\frac{3}{2}\right)-8\pi\Big], \quad ...
\end{align}
\begin{align}\label{eq:416}
\rho(\hat{r}) &= \sum_{n=0}^\infty c_n\hat{r}^n,\quad  
c_0=\hat{T}_\mathrm{c}^{3/2} \zeta \left(\textstyle\frac{3}{2}\right),
\nonumber \\
c_1 &=-2 \sqrt{\pi } a_2^{1/2} \hat{T}_\mathrm{c}^{3/2}
=-2 \hat{T}_\mathrm{c}^{7/4} \sqrt{\pi  
\zeta \left(\textstyle\frac{3}{2}\right)},
\nonumber \\
c_2 &= -\frac{\hat{T}_\mathrm{c}^{3/2} \left[a_2^{3/2} \zeta
   \left(\frac{1}{2}\right)+\sqrt{\pi } a_3\right]}{a_2^{1/2}}
   \nonumber \\
   &={\textstyle\frac{1}{9}}\hat{T}_\mathrm{c}^2 
   \left[8\pi -9\zeta \left(\textstyle\frac{1}{2}\right) \zeta
   \left(\textstyle\frac{3}{2}\right)\right],\quad \ldots
\end{align}



\end{document}